\documentclass{article}

\usepackage[table]{xcolor}
\usepackage{microtype}
\usepackage{graphicx}
\usepackage{subfigure}
\usepackage{booktabs} 

\usepackage{hyperref}

\usepackage[accepted]{icml2025}

\usepackage{amsmath}
\usepackage{amssymb}
\usepackage{mathtools}
\usepackage{amsthm}
\usepackage{multirow}
\usepackage{multicol}
\usepackage[export]{adjustbox}
\usepackage{caption}
\usepackage{amsmath}
\usepackage{bbding}
\usepackage{graphicx}
\usepackage{subfigure}
\usepackage{comment}
\usepackage{float}
\usepackage{wrapfig}
\usepackage{hyperref}       
\usepackage{url}            
\usepackage{booktabs}       
\usepackage{amsfonts}       
\usepackage{nicefrac}       
\usepackage{microtype}      

\usepackage[capitalize,noabbrev]{cleveref}

\theoremstyle{plain}

\theoremstyle{definition}

\theoremstyle{remark}

\usepackage[textsize=tiny]{todonotes}

\icmltitlerunning{Submission and Formatting Instructions for ICML 2025}

\begin{document}

\twocolumn[
\icmltitle{AffectGPT: A New Dataset, Model, and Benchmark for Emotion Understanding with Multimodal Large Language Models}




\begin{icmlauthorlist}
\icmlauthor{Zheng Lian}{1}
\icmlauthor{Haoyu Chen}{2}
\icmlauthor{Lan Chen}{1}
\icmlauthor{Haiyang Sun}{3}
\icmlauthor{Licai Sun}{2}
\icmlauthor{Yong Ren}{1}
\icmlauthor{Zebang Cheng}{4}
\icmlauthor{Bin Liu}{1}
\icmlauthor{Rui Liu}{5}
\icmlauthor{Xiaojiang Peng}{6}
\icmlauthor{Jiangyan Yi}{7}
\icmlauthor{Jianhua Tao}{7,8}
\end{icmlauthorlist}

\icmlaffiliation{1}{Institute of Automation, Chinese Academy of Sciences}
\icmlaffiliation{2}{CMVS, University of Oulu}
\icmlaffiliation{3}{Shanghai Jiao Tong University}
\icmlaffiliation{4}{Shenzhen University}
\icmlaffiliation{5}{Inner Mongolia University}
\icmlaffiliation{6}{Shenzhen Technology University}
\icmlaffiliation{7}{Department of Automation, Tsinghua University}
\icmlaffiliation{8}{Beijing National Research Center for Information Science and Technology, Tsinghua University}

\icmlcorrespondingauthor{Zheng Lian}{lianzheng2016@ia.ac.cn}
\icmlcorrespondingauthor{Jianhua Tao}{jhtao@tsinghua.edu.cn}

\icmlkeywords{Machine Learning, ICML}

\vskip 0.3in
]



\printAffiliationsAndNotice{}  

\begin{abstract}
    The emergence of multimodal large language models (MLLMs) advances multimodal emotion recognition (MER) to the next level—from naive discriminative tasks to complex emotion understanding with advanced video understanding abilities and natural language description. However, the current community suffers from a lack of large-scale datasets with intensive, descriptive emotion annotations, as well as a multimodal-centric framework to maximize the potential of MLLMs for emotion understanding. To address this, we establish a new benchmark for MLLM-based emotion understanding with a novel dataset (MER-Caption) and a new model (AffectGPT). Utilizing our model-based crowd-sourcing data collection strategy, we construct the largest descriptive emotion dataset to date (by far), featuring over 2K fine-grained emotion categories across 115K samples. We also introduce the AffectGPT model, designed with pre-fusion operations to enhance multimodal integration. Finally, we present MER-UniBench, a unified benchmark with evaluation metrics tailored for typical MER tasks and the free-form, natural language output style of MLLMs. Extensive experimental results show AffectGPT's robust performance across various MER tasks. We have released both the code and the dataset to advance research and development in emotion understanding: \href{https://github.com/zeroQiaoba/AffectGPT}{https://github.com/zeroQiaoba/AffectGPT}.
\end{abstract}

\section{Introduction}
Emotions encapsulate human intentions, and accurately recognizing emotional states is essential for enhancing human-computer interaction experiences \cite{minsky1988society}. Emotions can be conveyed through various human behaviors in different forms, giving rise to the task of multimodal emotion recognition (MER), which integrates multimodal information (e.g., audio, video, and text) to evaluate human emotional states. As a critical area in artificial intelligence, MER has broad applications, ranging from education \cite{schutz2007emotion} and psychological counseling \cite{liu2021towards} to empathic embodied robots \cite{spezialetti2020emotion}.

\begin{figure*}[t]
	\centering
	\includegraphics[width=0.93\linewidth]{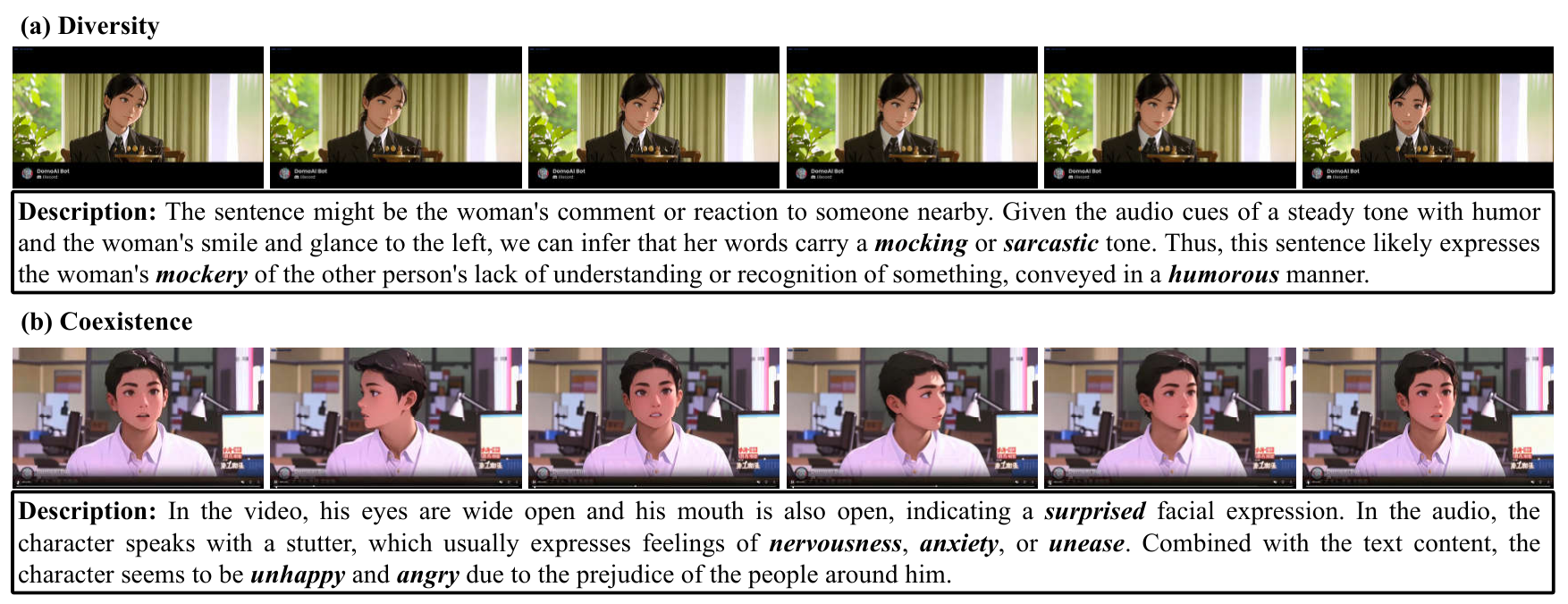}
	\caption{\textbf{Emotion complexity analysis}. Human emotions are often diverse and coexist simultaneously. Such complex emotional states are difficult to describe using discriminative frameworks. However, MLLMs can generate emotional descriptions, offering new possibilities for complex emotion modeling. Since the original videos contain real people, to address copyright concerns, we first use \href{https://www.domoai.app/zh-Hant/home}{DemoAI} to remove personal information and then proceed with visualization.}
	\label{Figure1}
\end{figure*}

Traditional methods primarily rely on discriminative models that map human emotions to the most likely categories from predefined emotion taxonomies. The most widely used taxonomy is Ekman's theory \cite{ekman1970universal}, which classifies all emotions into six basic categories: \emph{sadness}, \emph{happiness}, \emph{fear}, \emph{anger}, \emph{surprise}, and \emph{disgust}. However, such categorical frameworks exhibit some limitations in modeling human affective states. For example, our emotional expressions are diverse and nuanced due to culture-specific idioms \cite{matsumoto2001culture}, context-dependent metaphors \cite{kovecses2003metaphor}, and highly personalized behavioral patterns \cite{izard1993stability}. Current closed-set classification paradigms fail to capture the rich diversity of emotional expressions in real-world scenarios \cite{plutchik1980general}. Meanwhile, the rigid emotion taxonomies oversimplify the continuous spectrum of emotional experiences by forcing discrete labels (e.g., \emph{anger} or \emph{surprise}) onto nuanced affective states that often coexist \cite{cowen2017self}. Illustrations are provided in Figure \ref{Figure1}, where the diverse and coexisting issues are presented in Figs. (a) and (b).

Recent advances in multi-modal large language models (MLLMs) enable emotion understanding to move beyond traditional discriminative approaches, embracing a more generative framework \cite{liang2024survey}. This shift allows models to describe complex, coexisting emotional states in natural language. With the vast vocabulary, MLLMs can generate diverse, descriptive emotion categories beyond basic emotions, offering new opportunities for emotional understanding. However, recent research highlights that MLLMs still face limitations in emotion understanding \cite{lian2024open, lian2024gpt}. To address these challenges, this paper aims to advance emotional understanding from two key perspectives: the dataset and the model. Finally, we establish a unified benchmark tailored to the free-form, natural language output style of MLLMs.

\paragraph{\textbf{Dataset.}}
The current community still suffers from a lack of large-scale datasets with intensive, descriptive emotion annotations to realize the potential of MLLMs. The annotation strategies for constructing descriptive emotion datasets can be classified into three types: \emph{model-based}, \emph{human-based}, and \emph{human-model collaborative} strategies. The \emph{human-based} strategy is the most common way to construct emotion datasets with rich descriptive annotations. However, it's costly to conduct crowd-sourcing to scale up the dataset size with this purely manual annotation manner. Besides, humans tend to focus on main cues, resulting in brief and incomplete descriptions \cite{liu2022mafw}. Thus, researchers propose \emph{model-based} automatic annotation approaches. However, due to the lack of human proofreading, this approach may result in insufficient label quality \cite{cheng2024emotion}. Recently, \citet{lian2024open} propose a \emph{human-model collaborative} strategy, in which models provide pre-labeled cues and humans conduct multiple rounds of checks, which can be seen as a \emph{human-led, model-assisted} strategy. Although this approach offers more comprehensive descriptions, it is costly and difficult to scale the dataset. To balance label quality and dataset size, we introduce a novel annotation strategy that conducts model-based crowd-sourcing labeling with human priors, named \emph{model-led human-assisted}, to construct a large-scale emotion descriptive dataset with diverse emotional categories.

\begin{table*}[t]
	\centering
	\caption{\textbf{Dataset comparison.} ``I'', ``A'', ``V'', and ``T'' stand for image, audio, video, and text, respectively. We observe that descriptive datasets contain more diverse labels, providing the potential for modeling complex emotions.}
	\label{Table1}
	\scalebox{0.78}{
		\begin{tabular}{c|lrrccc}
			\hline
			&\textbf{Dataset} &\textbf{Modality} & \textbf{\# Samples} & \textbf{Description} &\textbf{\# Emotions} & {\begin{tabular}[c]{@{}c@{}}\textbf{Annotation} \textbf{Manner}\end{tabular}}   \\
			\hline
			\multirow{9}{*}{\begin{tabular}[c]{@{}c@{}}\textbf{Categorical}\\\textbf{Dataset}\end{tabular}} 
			& RAF-DB \cite{li2017reliable}        		   & I     & 29,672  & \textcolor{purple}{\XSolidBrush} & 7 & Human \\
			& AffectNet \cite{mollahosseini2017affectnet}  & I     & 450,000 & \textcolor{purple}{\XSolidBrush} & 8 & Human \\
			& EmoDB \cite{burkhardt2005database}     	   & A     & 535     & \textcolor{purple}{\XSolidBrush} & 7 & Human \\
			& MSP-Podcast \cite{lotfian2017building}       & A     & 73,042  & \textcolor{purple}{\XSolidBrush} & 8 & Human \\
			& DFEW \cite{jiang2020dfew}          		   & V     & 11,697  & \textcolor{purple}{\XSolidBrush} & 7 & Human \\ 
			& FERV39k \cite{wang2022ferv39k}       		   & V     & 38,935  & \textcolor{purple}{\XSolidBrush} & 7 & Human \\     
			& MER2023 \cite{lian2023mer}       		       & A,V,T & 5,030   & \textcolor{purple}{\XSolidBrush} & 6 & Human \\ 
			& MELD \cite{poria2019meld}          		   & A,V,T & 13,708  & \textcolor{purple}{\XSolidBrush} & 7 & Human \\ 
			\hline
			\multirow{8}{*}{\begin{tabular}[c]{@{}c@{}}\textbf{Descriptive}\\\textbf{Dataset}\end{tabular}}  
			& EmoVIT  \cite{xie2024emovit}       	   & I     & 51,200  & \textcolor{teal}{\Checkmark} & 988    & Model  \\
            & MERR-Coarse \cite{cheng2024emotion}      & A,V,T & 28,618  & \textcolor{teal}{\Checkmark} & 113    & Model  \\ 
			& MAFW \cite{liu2022mafw}          		   & A,V,T & 10,045  & \textcolor{teal}{\Checkmark} & 399    & Human \\
                & OV-MERD \cite{lian2024open}       	   & A,V,T & 332     & \textcolor{teal}{\Checkmark} & 236    & Human-led\scriptsize{+Model-assisted} \\ 
                & MERR-Fine \cite{cheng2024emotion}        & A,V,T & 4,487   & \textcolor{teal}{\Checkmark} & 484    & Human-led\scriptsize{+Model-assisted}  \\ 
			& \textbf{MER-Caption} 		   			   & A,V,T & 115,595 & \textcolor{teal}{\Checkmark} & 2,932  & Model-led\scriptsize{+Human-assisted} \\ 
			& \textbf{MER-Caption+}             & A,V,T & 31,327  & \textcolor{teal}{\Checkmark} & 1,972  & Model-led\scriptsize{+Human-assisted} \\ 
			\hline
		\end{tabular}
	}
\end{table*}

\paragraph{\textbf{Models.}} 
Existing MLLMs typically consist of three key components: a modality encoder that converts audio and video into low-level hidden features, a connector that transforms these features into a format more suitable for LLMs, and an LLM-based generator that produces responses based on the given instructions. While the results of MLLMs are promising, existing models generally leave everything of multimodal fusion to LLMs, which is insufficient for MER that emphasizes multimodal characteristics. This paper introduces the AffectGPT model, designed with a pre-fusion operation to emphasize multimodal integration.

\paragraph{\textbf{Benchmark.}}
Although it's desirable to generate emotional descriptions in a free-form, natural language style (see Appendix \ref{appendix:free_form}), this poses challenges for quantitative comparison. To address this, we propose metrics specifically designed for this output style. Additionally, to ensure fair and comprehensive evaluation, we introduce MER-UniBench, a benchmark that incorporates three typical tasks: fine-grained emotion recognition, basic emotion recognition, and sentiment analysis. We believe this work can enhance the emotion understanding capabilities of MLLMs and open possibilities for complex emotion modeling. The main contributions of this paper are summarized as follows:
\begin{itemize}
	\item We construct a large-scale emotional description dataset \textbf{MER-Caption}, which adopts a model-led, human-assisted annotation strategy to strike a balance between label quality and dataset size.
	
	\item We develop \textbf{AffectGPT}, which uses additional pre-fusion operations to enhance multimodal integration, thereby improving emotion understanding.
	
	\item We build \textbf{MER-UniBench}, which encompasses typical MER tasks with tailored metrics. This benchmark can offer comprehensive evaluation results for MLLM-based emotion understanding.

    \item Extensive experiments demonstrate the effectiveness of AffectGPT, which achieves over a 9\% performance improvement compared to existing MLLMs.
\end{itemize}

\section{MER-Caption: Dataset Construction}
Table \ref{Table1} summarizes existing emotion datasets, which can be broadly classified into categorical and descriptive datasets. The former directly provides emotion labels (e.g., \emph{happy}), while the latter offers textual descriptions related to emotions. We first conduct preliminary experiments to extract emotion labels from descriptive datasets (see Appendix \ref{appendix:label_extraction_prompt}). As shown in Table \ref{Table1}, descriptive datasets contain more diverse labels, offering the potential to capture complex emotions. Thus, this paper focuses on descriptive datasets.

\begin{figure*}[t]
	\centering
	\includegraphics[width=0.76\linewidth]{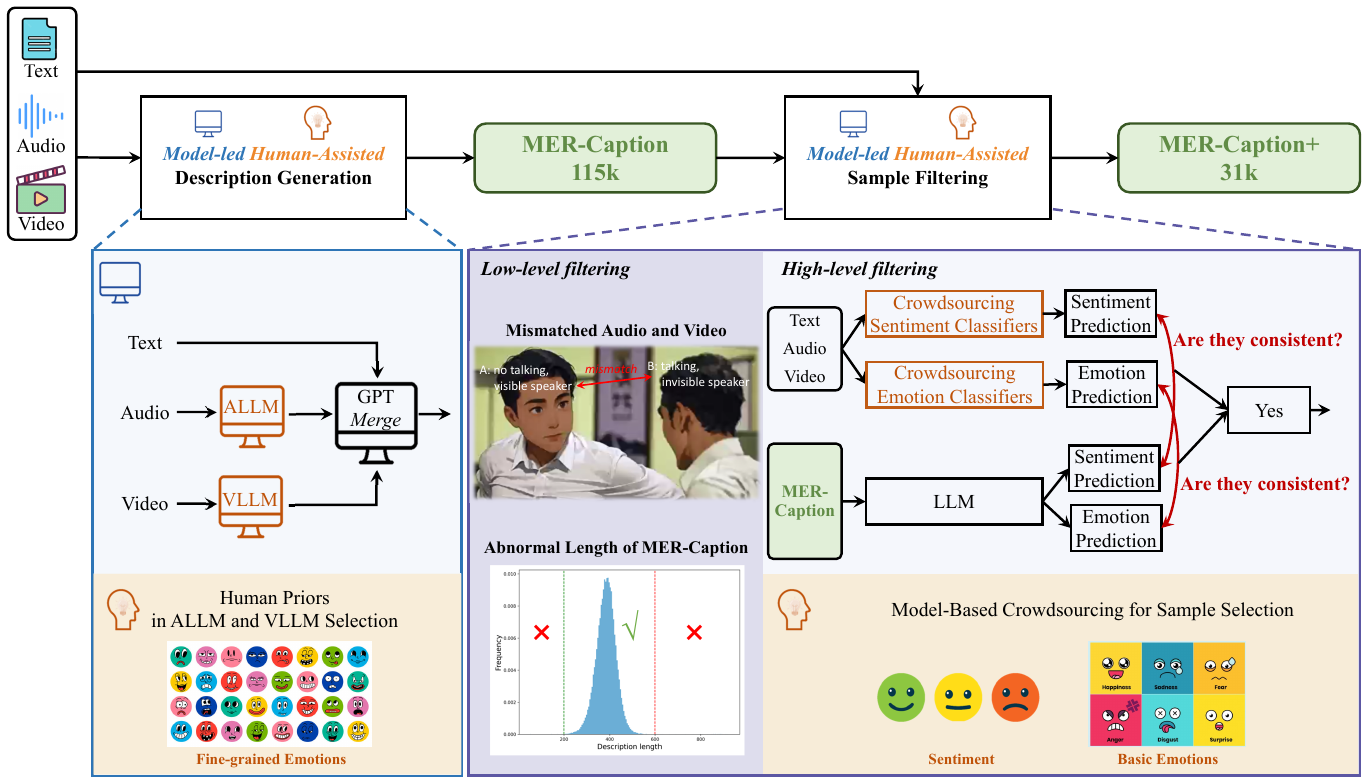}
	\caption{\textbf{Dataset construction pipeline.} To create a large-scale dataset with guaranteed label quality, we propose a \emph{model-led, human-assisted} annotation strategy. In this approach, we leverage human priors to guide description generation and sample filtering, ultimately achieving automatic annotation for unlabeled data.}
	\label{Figure2}
\end{figure*}

Based on the annotation manner, descriptive datasets can be categorized into \emph{model-based}, \emph{human-based}, and \emph{human-model collaborative} strategies (see Table \ref{Table1}). Although the model-based approach makes it easy to expand the dataset size, it mainly relies on experience to select models and lacks human intervention, resulting in insufficient label quality \cite{cheng2024emotion}. To enhance label quality, \citet{liu2022mafw} relied on human annotators to generate emotion descriptions. However, humans tend to focus on primary clues, easily leading to incomplete descriptions. To this end, \citet{lian2024open} proposed a \emph{human-led, model-assisted} strategy. Specifically, the model first provides pre-labeled descriptions, and then multiple annotators perform multi-round checks. Although this strategy produces more comprehensive descriptions, it comes with high annotation costs and faces challenges in scaling the dataset. In this paper, we review these annotation methods and introduce a \emph{model-led, human-assisted} strategy. As shown in Figure \ref{Figure2}, we leverage human priors to guide description generation and sample filtering, ultimately achieving automatic annotation for unlabeled data. Using this strategy, we construct the MER-Caption dataset, which includes 115K coarse-labeled samples and 31K fine-labeled samples, making a significant contribution to current descriptive datasets. The raw data in MER-Caption is sourced from the unlabeled portions of MER2024 \cite{lian2024mer}, with explicit permission from the dataset owners. Therefore, this paper does not involve the collection of new data but provides additional annotations for existing datasets. Appendix \ref{appendix:dataset_comparison} provides more comparisons with existing datasets.

\subsection{Description Generation}
\label{sec:human_assisted_model_selection}
The choice of base models is critical for generating accurate descriptions. Unlike previous work that relied solely on experience \cite{cheng2024emotion}, we guide model selection using \emph{human priors}. Specifically, we first select a small subset of samples for preliminary experiments. In this phase, we annotate fine-grained labels for each sample, allowing annotators to assign any emotions they deem appropriate, thus providing more diverse and precise labels. Based on the results in preliminary experiments (see Appendix \ref{appendix:multimodal_fusion}), we employ SALMONN \cite{tang2023salmonn} as the audio LLM (ALLM) to generate audio cues, Chat-UniVi \cite{jin2024chat} as the video LLM (VLLM) to extract visual cues, and GPT-3.5 \cite{openai2022chatgpt} (``gpt-3.5-turbo-16k-0613'') to merge the audio and video cues with text content. Then, to further reduce annotation costs, we experimented with replacing GPT-3.5 with other open-source LLMs but observed a drop in performance. The primary reason is that multimodal fusion in MER is inherently complex, often encountering issues such as modality conflict, where inconsistencies or contradictions arise between different modalities (see Appendix \ref{appendix:clue_merge_prompt}). This places high demands on the LLM's reasoning capabilities. Then, we adopt the above strategy for automatic annotation and create the MER-Caption dataset.

\subsection{Sample Filtering}
\label{sec:human_assisted_sample_filtering}
Since the descriptions generated by the above process have not been manually verified, MER-Caption inevitably contains some errors. To this end, we implement a two-level filtering process to enhance the label quality.

\paragraph{Low-level Filtering.}
First, we observe that some samples contain mismatched audio and video. As shown in Figure \ref{Figure2}, the visible person is not speaking, while the audio comes from an invisible person. This setup differs from our task, where we aim to analyze a person’s emotions based on their audio, video, and text content. Mismatched data complicates this task, shifting the focus to understanding how the interlocutor’s actions may influence the target person's emotions. Therefore, we remove such data and plan to address this issue in future work. To automatically determine whether the visible person is speaking, we use TalkNet \cite{tao2021someone}. Preliminary experiments indicate that this tool achieves over 90\% accuracy in identifying the speaking individual. Then, we remove samples with mismatched audio and video. Second, the length distribution of the generated descriptions roughly follows a Gaussian distribution (see Figure \ref{Figure2}). Preliminary experiments reveal that descriptions at both ends of the distribution are more likely to contain errors. For instance, when ALLM and VLLM (in Section \ref{sec:human_assisted_model_selection}) fail to generate responses, the resulting descriptions tend to be short. As a result, we further remove descriptions located at both ends of the distribution.

\paragraph{High-level Filtering.}
In addition to low-level filtering, we propose a \emph{model-based crowdsourcing} technique for high-level filtering. Specifically, we train multiple multimodal emotion and sentiment classifiers using human-annotated categorical datasets. Guided by MERBench \cite{lian2024merbench}, we use CLIP ViT-L \cite{radford2021learning} as the visual encoder and HUBERT-L \cite{hsu2021hubert} as the acoustic encoder, followed by an attention-based fusion strategy to make final emotion and sentiment predictions. These pre-trained models are then used to predict labels for unlabeled data, generating multiple predictions for each sample. To mitigate potential prediction errors, we apply majority voting to determine the final label, ensuring more reliable results. We refer to this process as \emph{model-based crowdsourcing}. Alternatively, emotions and sentiments can also be predicted based on the descriptions using the strategy outlined in Appendix \ref{appendix:label_extraction_prompt}. If the labels extracted from the descriptions differ from those obtained through \emph{model-based crowdsourcing}, we consider these descriptions to be of low quality and remove them. Through this process, we can extract knowledge from multiple human-based datasets to guide sample selection. After applying multi-level filtering, we obtain the \emph{MER-Caption+} dataset. Table \ref{Table1} presents detailed comparisons between our dataset and existing ones, highlighting that our dataset is the largest multimodal emotion description dataset with diverse emotion categories.

\begin{figure*}[t]
	\centering
	\includegraphics[width=\linewidth]{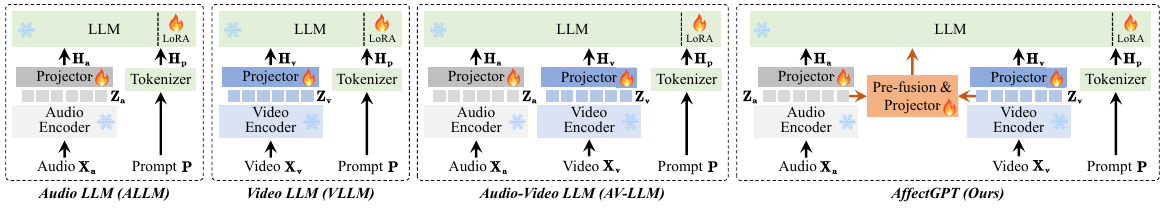}
	\caption{\textbf{Model comparison.} ALLM and VLLM primarily use modality-specific encoders and align them with the LLM through projection layers. AV-LLM mainly facilitates cross-modal interaction within the language model. In AffectGPT, we move the cross-modal interaction outside the language model and use a pre-fusion operation to enhance multimodal integration. In these figures, $\mathbf{P}$ can be determined based on the requirement of whether to include $\mathbf{X_t}$.}
	\label{Figure3}
\end{figure*}

\section{AffectGPT: Model Design}
Our primary goal is to map audio-video-text inputs to emotion-related descriptions. In this section, we first review the current mainstream architectures. We then introduce AffectGPT, a model specifically designed to highlight multimodal characteristics in emotion understanding.

\subsection{Mainstream Architecture}
MLLM aims to understand multimodal input and generate appropriate responses based on the input and user instructions. Unlike pure-text LLMs, the primary challenge for MLLMs lies in enabling the model to perceive multimodal input, i.e., providing the model with ``eyes'' and ``ears''. In existing models, the most common approach is to first extract modality-specific embeddings and then align them with the LLM through projection layers. For audio-video joint tasks, Audio-Video LLMs (AV-LLMs) typically facilitate cross-modal interaction within the language model. Figure \ref{Figure3} illustrates the current mainstream architecture.

Formally, for each sample $\mathbf{X}$, we represent its video, audio, and text content as $\mathbf{X_v}$, $\mathbf{X_a}$, and $\mathbf{X_t}$, respectively. Given an instruction $\mathbf{Q}$, the goal is to output the correct response $\mathbf{R}$. For the visual input $\mathbf{X_v}$, we use a video expert to encode it into a latent space $\mathbf{Z_v}$, then apply a projector $G_v(\cdot)$ to generate visual tokens $\mathbf{H_v}=G_v(\mathbf{Z_v})$. Similarly, for the acoustics input $\mathbf{X_a}$, we use an audio expert and a projector to generate the acoustic embeddings $\mathbf{Z_a}$ and tokens $\mathbf{H_a}$. For the instruction $\mathbf{Q}$ and text content $\mathbf{X_t}$, we use a template to merge them into a prompt $\mathbf{P}$, and then map them to the corresponding tokens through the tokenizer and embedding layer in the language model. After obtaining these tokens, we concatenate them and feed them into the LLM decoder. The primary objective is to maximize the likelihood of the target response $\mathbf{R}$, conditioned on multimodal content ($\mathbf{X_v}$, $\mathbf{X_a}$, $\mathbf{X_t}$) and user instruction $\mathbf{Q}$:
\begin{equation}
\label{eq:loss_function}
\max P\left(\mathbf{R}|\mathbf{X_v}, \mathbf{X_a}, \mathbf{X_t}, \mathbf{Q}\right).
\end{equation}
The above formula is optimized in an autoregressive manner, consistent with the objective function of LLMs. We represent the response as $\mathbf{R}=\{r_i\}_{i=1}^{L_r}$, where $L_r$ is the number of tokens. Then, Eq. \ref{eq:loss_function} is transformed into:
\begin{equation}
\max \prod_{l=1}^{L_r} P\left(r_l|\mathbf{X_v}, \mathbf{X_a}, \mathbf{X_t}, \mathbf{Q}, \mathbf{R}_{<l}\right).
\end{equation}
In this equation, $r_l$ is the current token to be predicted, and $\mathbf{R}_{<l}=\{r_i\}_{i=1}^{l-1}$ is the previously generated tokens, which serve as additional conditioning during training.

\subsection{Pre-fusion Operation}
Mainstream AV-LLMs leave everything of cross-modal interaction to the LLMs (in Figure \ref{Figure3}), which is insufficient for handling MER with multimodal characteristics. To address this, we propose a pre-fusion operation that moves the cross-modal interaction outside the LLMs, further enhancing multimodal integration. We refer to this model as AffectGPT. This paper introduces two types of pre-fusion operations: Q-Former-based and attention-based pre-fusion. By default, we apply this operation to $\mathbf{Z_v} \in \mathbb{R}^{t_v \times d}$ and $\mathbf{Z_a} \in \mathbb{R}^{t_a \times d}$. We also experimented with $\mathbf{H_v}$ and $\mathbf{H_a}$, but this choice led to a decrease in performance.

\paragraph{Q-Former.}
In this module, we preserve the temporal information in the vision features $\mathbf{Z_v}$ and audio features $\mathbf{Z_a}$, and utilize Q-Former \cite{li2023blip} for multimodal fusion. Specifically, to compress the multimodal content, we first create $K$ learnable query tokens $\mathbf{Z_q} \in \mathbb{R}^{K \times d}$. Then, we interact $\mathbf{Z_q}$ with the concatenated $\mathbf{Z_v}$ and $\mathbf{Z_a}$ through cross-attention, thereby distilling the knowledge from the multimodal content into the query tokens. Formally, this process can be represented as:
\begin{equation}
\mathbf{Z_{av}} = \mbox{Concat}\left(\mathbf{Z_a}, \mathbf{Z_v}\right),
\end{equation}
\begin{equation}
\mathbf{Z_f} = \mbox{Q-Former}\left(\mathbf{Z_q}, \mathbf{Z_{av}} + \mbox{PE}\left(\mathbf{Z_{av}}\right)\right),
\end{equation}
where $\mathbf{Z_{av}} \in \mathbb{R}^{(t_a+t_v) \times d}$, with the concatenation operation applied along the temporal dimension. Here, $\mathbf{Z_f} \in \mathbb{R}^{K \times d}$, and $\mbox{PE}(\cdot)$ represents the positional encoding.

\paragraph{Attention.}
Unlike Q-Former which preserves temporal information, we propose a simpler architecture that directly compresses temporal information and applies attention mechanisms for multimodal fusion. This simplified module is inspired by MERBench \cite{lian2024merbench}, which proves that in MER tasks, features with temporal information do not always lead to better performance than compressed features. Formally, we first apply average pooling to compress unimodal features. Then, we calculate the attention weights to emphasize important modalities:
\begin{equation}
\mathbf{\hat{Z}_a} = \mbox{Pooling}\left(\mathbf{Z}_a\right), \mathbf{\hat{Z}_v} = \mbox{Pooling}\left(\mathbf{Z}_v\right),
\end{equation}
\begin{equation}
\mathbf{\hat{Z}_{av}} = \mbox{Concat}\left(\mathbf{\hat{Z}_a}, \mathbf{\hat{Z}_v}\right),
\end{equation}
\begin{equation}
\mathbf{Z_f} = \mathbf{\hat{Z}_{av}}^T \left(\mathbf{W} \cdot \mbox{Flatten}\left(\mathbf{\hat{Z}_{av}}\right)\right),
\end{equation}
where $\mathbf{\hat{Z}_a} \in \mathbb{R}^{d}$, $\mathbf{\hat{Z}_v} \in \mathbb{R}^{d}$, $\mathbf{\hat{Z}_{av}} \in \mathbb{R}^{2 \times d}$, and $\mathbf{W} \in \mathbb{R}^{2 \times 2d}$. Finally, we obtain the fused features $\mathbf{Z_f} \in \mathbb{R}^{d}$.

Regarding computational efficiency, the pre-fusion operation relies on Q-Former or attention mechanisms, which are significantly less computationally intensive than LLMs. Theoretically, the Q-Former enables cross-modal interaction by distilling multimodal content into query tokens, whereas the attention mechanism achieves this by dynamically computing attention weights based on multimodal inputs.

\section{MER-UniBench: Evaluation Benchmark}
\label{sec:mer_unibench}
We introduce MER-UniBench, a comprehensive evaluation benchmark designed to cover typical MER tasks. Given the free-form, natural language output style of MLLMs (see Appendix \ref{appendix:free_form}), we also design specialized evaluation metrics. More details can be found in Appendix \ref{appendix:dataset_details}.

\begin{table*}[t]
	\centering
	\renewcommand\tabcolsep{5.6pt}
	\renewcommand\arraystretch{1.06}
	\caption{\textbf{Main results.} This table presents the results for the primary metrics, with Section \ref{sec:mer_unibench} outlining the primary metrics for each task. The values for other metrics can be found in Appendix \ref{appendix:complte_results}. In this table, ``MOSI'', ``MOSEI'', ``SIMS'', and ``SIMS v2'' refer to CMU-MOSI, CMU-MOSEI, CH-SIMS, and CH-SIMS v2, respectively. The last column shows the dataset-wise mean score, i.e., the average score across all datasets.}
	\label{Table2}
	\scalebox{0.8}{
		\begin{tabular}{lccc|cccc|cccc|c|c}
			\hline
			&\multicolumn{3}{c|}{\textbf{Modality}} 
                &\multicolumn{4}{c|}{\textbf{Basic}} 
                &\multicolumn{4}{c|}{\textbf{Sentiment}} 
                &{\textbf{Fine-grained}} &\multirow{2}{*}{\textbf{Mean}} \\
                & A & V & T & MER2023 & MER2024 & MELD &IEMOCAP &MOSI &MOSEI &SIMS &SIMS v2 & OV-MERD+ & \\
			\hline
            
            OneLLM        &$\surd$  &$\times$ &$\surd$ & 25.52 & 17.21 & 28.32 & 33.44 & 64.01 & 54.09 & 63.39 & 61.98 & 22.25 & 41.14\\
SECap         &$\surd$  &$\times$ &$\surd$ & 40.95 & 52.46 & 25.56 & 36.92 & 55.76 & 54.18 & 59.51 & 57.41 & 36.97 & 46.64\\
PandaGPT      &$\surd$  &$\times$ &$\surd$ & 33.57 & 39.04 & 31.91 & 36.55 & 66.06 & 61.33 & 62.93 & 58.88 & 31.33 & 46.84\\
Qwen-Audio    &$\surd$  &$\times$ &$\surd$ & 41.85 & 31.61 &49.09& 35.47 & 70.09 & 46.90 &70.73& 65.26 & 32.36 & 49.26\\
SALMONN       &$\surd$  &$\times$ &$\surd$ & 55.53 & 45.38 & 45.62 & 46.84 &81.00& 67.03 & 68.69 & 65.93 & 45.00 & 57.89\\
\textbf{AffectGPT}     &$\surd$ &$\times$ &$\surd$ &\textbf{72.94}&\textbf{73.41}&\textbf{56.63}&\textbf{55.68}&\textbf{83.46}&\textbf{80.74}&\textbf{82.99}&\textbf{83.75}&\textbf{59.98}&\textbf{72.18}\\
            
            \hline

            Otter         &$\times$ &$\surd$  &$\surd$ & 16.41 & 14.65 & 22.57 & 29.08 & 52.89 & 50.44 & 57.56 & 53.12 & 16.63 & 34.82\\
Video-LLaVA   &$\times$ &$\surd$  &$\surd$ & 36.93 & 30.25 & 30.73 & 38.95 & 56.37 & 61.64 & 53.28 & 57.45 & 34.00 & 44.40\\
PandaGPT      &$\times$ &$\surd$  &$\surd$ & 39.13 & 47.16 & 38.33 & 47.21 & 58.50 & 64.25 & 62.07 & 65.25 & 35.07 & 50.77\\
Video-ChatGPT &$\times$ &$\surd$  &$\surd$ & 44.86 & 46.80 & 37.33 &56.83& 54.42 & 63.12 & 64.82 & 65.80 & 39.80 & 52.64\\
VideoChat2    &$\times$ &$\surd$  &$\surd$ & 33.67 & 54.50 & 36.64 & 48.70 & 66.84 & 54.32 & 69.49 & 70.66 & 39.21 & 52.67\\
LLaMA-VID     &$\times$ &$\surd$  &$\surd$ & 50.72 & 57.60 & 42.75 & 46.02 & 61.78 & 63.89 & 69.35 & 67.48 & 45.01 & 56.07\\
VideoChat     &$\times$ &$\surd$  &$\surd$ & 48.73 & 57.30 & 41.11 & 48.38 & 65.13 & 63.61 & 69.52 &72.14& 44.52 & 56.71\\
Chat-UniVi    &$\times$ &$\surd$  &$\surd$ &57.62&65.67& 45.61 & 52.37 & 54.53 & 63.18 & 68.15 & 66.36 &48.00&57.94\\
mPLUG-Owl     &$\times$ &$\surd$  &$\surd$ &56.86&59.89&49.11&55.54&72.40&72.91&72.13&75.00&48.18&62.45\\
\textbf{AffectGPT}     &$\times$ &$\surd$ &$\surd$ &\textbf{74.58}&\textbf{75.29}&\textbf{57.63}&\textbf{62.19}&\textbf{82.39}&\textbf{81.57}&\textbf{87.20}&\textbf{86.29}&\textbf{61.65}&\textbf{74.31}\\
			
            \hline
            
            PandaGPT      &$\surd$  &$\surd$  &$\surd$ & 40.21 & 51.89 & 37.88 & 44.04 & 61.92 &67.61& 68.38 & 67.23 & 37.12 & 52.92\\
Emotion-LLaMA &$\surd$ &$\surd$ &$\surd$ & 59.38 & 73.62 & 46.76 & 55.47 & 66.13 & 67.66 & 78.32 & 77.23 & 52.97 & 64.17 \\
\textbf{AffectGPT}     &$\surd$ &$\surd$ &$\surd$ &\textbf{78.54}&\textbf{78.80}&\textbf{55.65}&\textbf{60.54}&\textbf{81.30}&\textbf{80.90}&\textbf{88.49}&\textbf{86.18}&\textbf{62.52}&\textbf{74.77}\\
			\hline
            
		\end{tabular}
	}
\end{table*}

\paragraph{Fine-grained Emotion Recognition.}
This task enables the prediction of fine-grained emotions, extending beyond basic categories. OV-MERD \cite{lian2024open} is a typical dataset for this task. To improve the reliability of the evaluation results, we expand its dataset size, referring to it as OV-MERD+. For the evaluation metrics, we draw inspiration from previous work \cite{lian2024open} and calculate results in two steps: eliminating the impact of synonyms and using set-level metrics. First, we apply a three-level grouping strategy to mitigate the impact of synonyms:
\begin{itemize}
	\item \textbf{Level 1.} We map different forms of emotion words to their base form. For example, we map \emph{happier} and \emph{happiness} to \emph{happy}. This function is denoted as $F_{l_1}(\cdot)$.
	
	\item \textbf{Level 2.} We map synonyms to a unified label. For example, we map \emph{happy} and \emph{joyful} to \emph{happy}. This mapping function is represented as $F_{l_2}(\cdot)$.
	
	\item \textbf{Level 3.} Emotion wheel provides natural grouping information, with core emotions displayed in the inner part and more nuanced labels in the outer part \cite{plutchik1980general}. Since there is no consensus on the emotion wheel, we use $K$ emotion wheels (see Appendix \ref{appendix:emotion_wheel}). For each sector of the emotion wheel $w_k, k \in [1, K]$, we map all outer labels to the corresponding inner labels. This mapping function is denoted as $F_{l_3}^{w_k}(\cdot)$.
\end{itemize}

The above grouping functions can be summarized as:
\begin{equation}
G_{w_k}(\cdot) = F_{l_3}^{w_k}{\left(F_{l_2}\left(F_{l_1}\left(\cdot\right)\right)\right)}, k \in [1, K].
\end{equation}
For each sample, the number of labels is variable. Therefore, we define a set-based evaluation metric. Specifically, suppose the dataset contains $N$ samples. For sample $x_i$, the true labels are $\mathbf{Y}_i=\{y_i^j\}_{j=1}^{n_i}$, and the predicted labels are $\mathbf{\hat{Y}}_i=\{\hat{y}_i^j\}_{j=1}^{\hat{n}_i}$. The evaluation metric is defined as follows:
\begin{equation}
\mbox{Precision}_{\mbox{s}}^{k} = \frac{1}{N}\sum_{i=1}^{N}\frac{\left|G_{w_k}( \mathbf{Y}_i ) \cap G_{w_k}(\mathbf{\hat{Y}}_i)\right|}{\left|G_{w_k}(\mathbf{\hat{Y}}_i)\right|},
\end{equation}
\begin{equation}
\mbox{Recall}_{\mbox{s}}^{k} = \frac{1}{N}\sum_{i=1}^{N}\frac{\left|G_{w_k}( \mathbf{Y}_i ) \cap G_{w_k}(\mathbf{\hat{Y}}_i)\right|}{\left|G_{w_k}(\mathbf{{Y}}_i)\right|},
\end{equation}
\begin{equation}
\mbox{F}_{\mbox{s}}^{k} = 2\times\frac{\mbox{Precision}_{\mbox{s}}^{k}\times\mbox{Recall}_{\mbox{s}}^{k}}{\mbox{Precision}_{\mbox{s}}^{k}+\mbox{Recall}_{\mbox{s}}^{k}}.
\end{equation}
Finally, we compute the average results across different emotion wheels for ranking. Take $\mbox{F}_{\mbox{s}}$ as an example:
\begin{equation}
\mbox{F}_{\mbox{s}} = \frac{1}{K}\sum_{k=1}^{K}\mbox{F}_{\mbox{s}}^{k}.
\end{equation}
Here, $\mbox{Precision}_{\mbox{s}}$ indicates the number of correctly predicted labels, and $\mbox{Recall}_{\mbox{s}}$ indicates whether the prediction covers all ground truth. $\mbox{F}_{\mbox{s}}$ is a harmonic mean of two metrics. Since $\mbox{F}_{\mbox{s}}$ considers both accuracy and completeness, we use it as the primary metric, with $\mbox{Precision}_{\mbox{s}}$ and $\mbox{Recall}_{\mbox{s}}$ serving as secondary metrics. To extract the predicted emotions $\mathbf{\hat{Y}}_i$, we employ the strategy mentioned in Appendix \ref{appendix:label_extraction_prompt}.

\paragraph{Basic Emotion Recognition.}
This task is a key branch of MER, whose main goal is to select the most likely label from a fixed set of basic emotions. For this task, we select four widely used benchmark datasets: MER2023 \cite{lian2023mer}, MER2024 \cite{lian2024mer}, IEMOCAP \cite{busso2008iemocap}, and MELD \cite{poria2019meld}. However, the output of MLLMs, $\mathbf{\hat{Y}}_i=\{\hat{y}_i^j\}_{j=1}^{\hat{n}_i}$, contains a variable number of labels, while the dataset only provides one true label $y_i$. In this case, traditional metrics (such as \emph{accuracy}) are not suitable for performance evaluation. To address this, we propose a new metric, \emph{hit rate}, which is set to 1 when $y_i \in \mathbf{\hat{Y}}_i$ and 0 otherwise. Considering that $\mathbf{\hat{Y}}_i$ is in free-form and $y_i$ belongs to basic emotions $\mathcal{Y}$, we may encounter cases where $\hat{y}_i \notin \mathcal{Y}$. To this end, we use the mapping function $G_{w_k}(\cdot)$ and define the metric as follows:
\begin{equation}
\mbox{HIT} = \frac{1}{N}\sum_{i=1}^{N} \mathbb{I}\left[G_{w_k}(y_i) \in G_{w_k}(\mathbf{\hat{Y}}_i) \right],
\end{equation}
where $\mathbb{I}[\cdot]$ is an indicator function. The motivation for this metric stems from the fact that basic emotion recognition tasks typically provide majority-voted labels $y_i$, which are generally reliable. However, emotion descriptions produce free-form outputs $\mathbf{\hat{Y}}_i$ that may contain multiple labels, including fine-grained ones beyond basic emotions. Therefore, we use the \emph{hit rate} as the metric, ensuring that the basic label $y_i$ should be at least in $\mathbf{\hat{Y}}_i$.

During the design of this metric, we also explored the possibility of evaluating potentially incorrect labels in $\mathbf{\hat{Y}}_i$. However, the labels in $\mathbf{\hat{Y}}_i$ that differ from the basic label $y_i$ are not necessarily incorrect - they may represent some fine-grained emotions not covered by basic categories. Since basic emotion recognition tasks lack fine-grained reference labels, we have not yet established appropriate evaluation metrics for this purpose. This remains an important research direction for our future work.

\paragraph{Sentiment Analysis.}
This task is more fundamental than the two tasks mentioned above, aiming to predict the sentiment polarity. For this task, we select four benchmark datasets: CMU-MOSI \cite{zadeh2017tensor}, CMU-MOSEI \cite{zadeh2018multimodal}, CH-SIMS \cite{yu2020ch}, and CH-SIMS v2 \cite{liu2022make}. For these benchmark datasets, the original labels are floating-point values, ranging from $[-1, 1]$ or $[-3, 3]$. We map scores of $<0$ to negative sentiment and scores of $>0$ to positive sentiment. To extract sentiment labels from the MLLM's output, we employ the strategy outlined in Appendix \ref{appendix:label_extraction_prompt}. Following previous work \cite{zadeh2017tensor, zadeh2018multimodal}, we evaluate performance using accuracy (ACC) and weighted average F-score (WAF). Due to the inherent label imbalance, we choose WAF as the primary metric and ACC as the secondary metric.

\section{Results and Discussion}
\label{sec:results}
In this section, we present the experimental results and provide an in-depth analysis. Detailed implementation information can be found in Appendix \ref{appendix:implementation_details}.

\paragraph{Main Results.}
We compare the performance of AffectGPT with other MLLMs on MER-UniBench. Since our inputs include audio, video, and text content, we only select MLLMs that support at least audio or video. For models that support both audio and video, we test different modality combinations. Model cards are provided in Appendix \ref{appendix_sec:mllm}. To ensure a fair comparison, we use their official weights and input corresponding multimodal content, asking them to infer the emotional state. In Table \ref{Table2}, AffectGPT significantly outperforms existing MLLMs. This can be attributed to the fact that current instruction datasets pay little attention to MER tasks. Additionally, existing models place the entire multimodal fusion within the LLM, which is insufficient for MER tasks that require effective multimodal integration. By leveraging our newly proposed dataset and model, we provide a promising approach to enhancing emotion understanding capability in MLLMs. Meanwhile, for different datasets, increasing the input modality does not always improve performance, as it may also introduce irrelevant information that interferes with emotional understanding.

\begin{table}[t]
	\centering
	\caption{\textbf{Dataset comparison}. We only change the training dataset, keeping all other aspects consistent. This table reports the mean score across all datasets in MER-UniBench.}
	\label{Table3}
	\scalebox{0.8}{
		\begin{tabular}{c|lc|c}
			\hline
			& \textbf{Dataset} 
			& \textbf{Filtering}
			& \textbf{MER-UniBench} \\
			\hline	
			\multirow{8}{*}{\begin{tabular}[c|]{@{}c@{}}\textbf{General}\\\textbf{Instruction}\end{tabular}} 
			& \multirow{2}{*}{MiniGPT4} 
			& $\times$  & 31.74 \\
			& & $\surd$ & 35.53\\
			\cline{2-4}
			& \multirow{2}{*}{VideoChat} 
			& $\times$  & 37.16 \\
			& & $\surd$ & 37.63 \\
			\cline{2-4}
			& \multirow{2}{*}{LLaVA} 
			& $\times$   & 46.69 \\
			& & $\surd$  & 46.27 \\
			\cline{2-4}
			& \multirow{2}{*}{WavCaps} 
			& $\times$ & 21.65 \\
			& & $\surd$ & 37.91 \\
			\hline
			\multirow{6}{*}{\begin{tabular}[c|]{@{}c@{}}\textbf{Emotion}\\\textbf{Description}\end{tabular}} 
			& EmoVIT      & -- & 51.05 \\
			& MAFW        & -- & 58.16 \\
			& MERR-Coarse & -- & 49.85 \\
			& MERR-Fine   & -- & 64.55 \\
			& MER-Caption & -- & 68.91 \\
			& MER-Caption+ & -- & \textbf{74.77} \\
			\hline
		\end{tabular}
	}
\end{table}

\paragraph{Effectiveness of MER-Caption.}
\label{sec:ablation_study_on_mercaption}
Table \ref{Table3} compares the performance of MER-Caption with existing datasets. For a fair comparison, we use the same model architecture and experimental settings and only change the training data. For general instruction datasets, we further conduct filtering experiments to remove samples without emotion-related content, emphasizing emotion-related subsets. Specifically, we use the prompt in Appendix \ref{appendix:label_extraction_prompt} and extract emotion labels from each instruction-answer pair. Samples yielding empty emotion outputs are removed.

In Table \ref{Table3}, the excellent performance of MER-Caption proves the limitations of current datasets in addressing MER. On the one hand, general instruction datasets pay insufficient attention to emotion-related tasks. On the other hand, emotional description datasets often suffer from inadequate dataset scales or insufficient annotation quality. Therefore, our dataset can serve as an important complement to existing datasets. Meanwhile, for the general instruction datasets, the filtering approach is less effective on the LLaVA and VideoChat datasets. We hypothesize that the detailed descriptions in non-emotion subsets may also provide valuable cues for inferring emotional states in some scenarios. 

Furthermore, we would like to acknowledge that MER-Caption+ may contain inaccurate descriptions due to the use of an automatic annotation strategy without manual checks. However, the experimental results in Table \ref{Table3} show that MER-Caption+ achieves significantly better performance than the manually annotated MAFW dataset. The main reason is that humans tend to focus on major clues, which can easily lead to incomplete descriptions. These results confirm that, despite the lack of manual checks in MER-Caption+, we can still ensure the quality of the labels. In the future, we will investigate other post-filtering techniques to further improve MER-Caption+'s annotation quality.

\paragraph{Ablation Study on MER-Caption.}
As shown in Table \ref{Table4}, compared to the results without filtering or with only low-level filtering, our two-level filtering leads to a performance improvement, further verifying the effectiveness of our filtering technique. These findings underscore that dataset quality is as critical as quantity, and fewer training samples do not necessarily lead to worse performance. Please see Appendix \ref{appendix:ablation_study_dataset} for more details.

\begin{table}[t]
	\centering
	\caption{\textbf{Necessity of filtering.} Besides the results on MER-UniBench, we also provide task-level results. ``E'' and ``S'' are abbreviations for emotion and sentiment, respectively.}
	\label{Table4}
	\scalebox{0.8}{
		\begin{tabular}{cc|ccc|c}
			\hline
			\textbf{Low} & \textbf{High} & \textbf{Fine-grained} & \textbf{Basic} & \textbf{Sentiment} & \textbf{MER-UniBench} \\
			\hline
			$\times$  & $\times$    & 58.42 & 64.36 & 76.09 & 68.91\\
			$\surd$   & $\times$    & 58.61 & 62.78 & 79.04 & 69.54\\
			$\surd$   & E 		& 61.72 & 67.83 & 82.74 & 73.78\\
			$\surd$   & S 		& 61.00 & 66.33 & \textbf{85.04} & 74.05\\
			$\surd$   & E+S  & \textbf{62.52} & \textbf{68.38} & 84.22 & \textbf{74.77}\\ 
			\hline
		\end{tabular}
	}
\end{table}

\begin{table}[t]
	\centering
	\renewcommand\tabcolsep{5pt}
	\renewcommand\arraystretch{1.06}
	\caption{{Role of pre-fusion operation.}}
	\label{Table5}
	\scalebox{0.8}{
		\begin{tabular}{c|ccc|c}
			\hline
			\textbf{Pre-fusion} & \textbf{Fine-grained} & \textbf{Basic} & \textbf{Sentiment} & \textbf{MER-UniBench} \\
			\hline	
			$\times$  & 61.21 & 66.28 & 82.57  & 72.95 \\
			\hline
			Q-Former  & \textbf{62.65} & 66.89 & \textbf{84.30}  & 74.16 \\
			Attention & {62.52} & \textbf{68.38} & 84.22 & \textbf{74.77}\\
			\hline
		\end{tabular}
	}
\end{table}

\paragraph{Ablation Study on Model.}
\label{sec:architecture}
Table \ref{Table5} compares different architectures and examines the impact of pre-fusion operations. Our results show that pre-fusion operations generally improve performance. This highlights the importance of treating cross-modal interactions as separate modules to more effectively capture multimodal characteristics. 

\paragraph{Analysis of Input Impact.}
Table \ref{Table20} reveals the impact of different inputs. The distinction between ``face'' and ``frame'' lies in whether an additional face extractor is used to extract faces from frames. We observe a general trend: multimodal results outperform unimodal results. These findings suggest that humans express emotions through multiple modalities, and integrating them leads to improved performance. Additionally, face inputs slightly outperform frame inputs, and their combination does not result in further improvement. This suggests that current MER datasets mainly focus on people, with limited emotional information conveyed through the environment. As a result, in this paper, we default to using audio, face, and text as the inputs.

\paragraph{User Study.}
We conduct a user study to evaluate the quality of our proposed dataset. Since MERR-Fine and MERR-Coarse \cite{cheng2024emotion} share some samples with our dataset, we randomly select 20 overlapping samples. We then hire four expert annotators and present them with two descriptions for each sample: one from our dataset and one from the other datasets. The annotators are asked to watch the video and select the more accurate description. As shown in Table \ref{Table15}, our dataset provides more accurate descriptions than both the model-based MERR-Coarse and the human-filtered MERR-Fine, thereby validating the effectiveness of our proposed annotation strategy.

\begin{table}[!t]
	\centering
	\caption{\textbf{Input impact analysis.} The difference between ``face'' and ``frame'' is whether an additional face extractor is used to extract faces from frames.}
	\label{Table20}
	\scalebox{0.8}{
		\begin{tabular}{l|cccc|c}
			\hline
			& \multicolumn{4}{c|}{\textbf{Input}}  & \multirow{2}{*}{\textbf{MER-UniBench}} \\
			& audio& face &frame & text & \\
			\hline
			\multirow{4}{*}{\textbf{unimodal}} & $\surd$  & $\times$ & $\times$ & $\times$ & 60.08 \\
			& $\times$ & $\surd$  & $\times$ & $\times$ & 60.47 \\
			& $\times$ & $\times$ & $\surd$  & $\times$ & 59.47 \\
			& $\times$ & $\times$ & $\times$ & $\surd$  & 67.44 \\
                \hline
			\multirow{3}{*}{\textbf{multimodal}} 
            & $\surd$  & $\surd$  & $\times$ & $\surd$  & 74.77 \\
            & $\surd$  & $\times$ & $\surd$  & $\surd$  & 73.39 \\
			& $\surd$  & $\surd$  & $\surd$  & $\surd$  & 74.60 \\
			\hline
		\end{tabular}
	}
\end{table}

\begin{table}[t]
	\centering
	\caption{User study.}
	\label{Table15}
	\scalebox{0.8}{
		\begin{tabular}{l|ccc}
			\hline
			\multirow{2}{*}{\textbf{Dataset}} & \multicolumn{3}{c}{\textbf{MER-Caption+}}  \\
			& Wins	& Losses & Ties \\
			\hline
			MERR-Coarse & 0.86 & 0.04 & 0.10 \\
                MERR-Fine   & 0.59 & 0.27 & 0.14 \\
			\hline
		\end{tabular}
	}
\end{table}

\begin{figure}[t]
	\begin{center}
		\subfigure[\scriptsize{LLM}]{
			\label{Figure60-1}
			\centering
			\includegraphics[width=0.28\linewidth, trim=0 0 0 0]{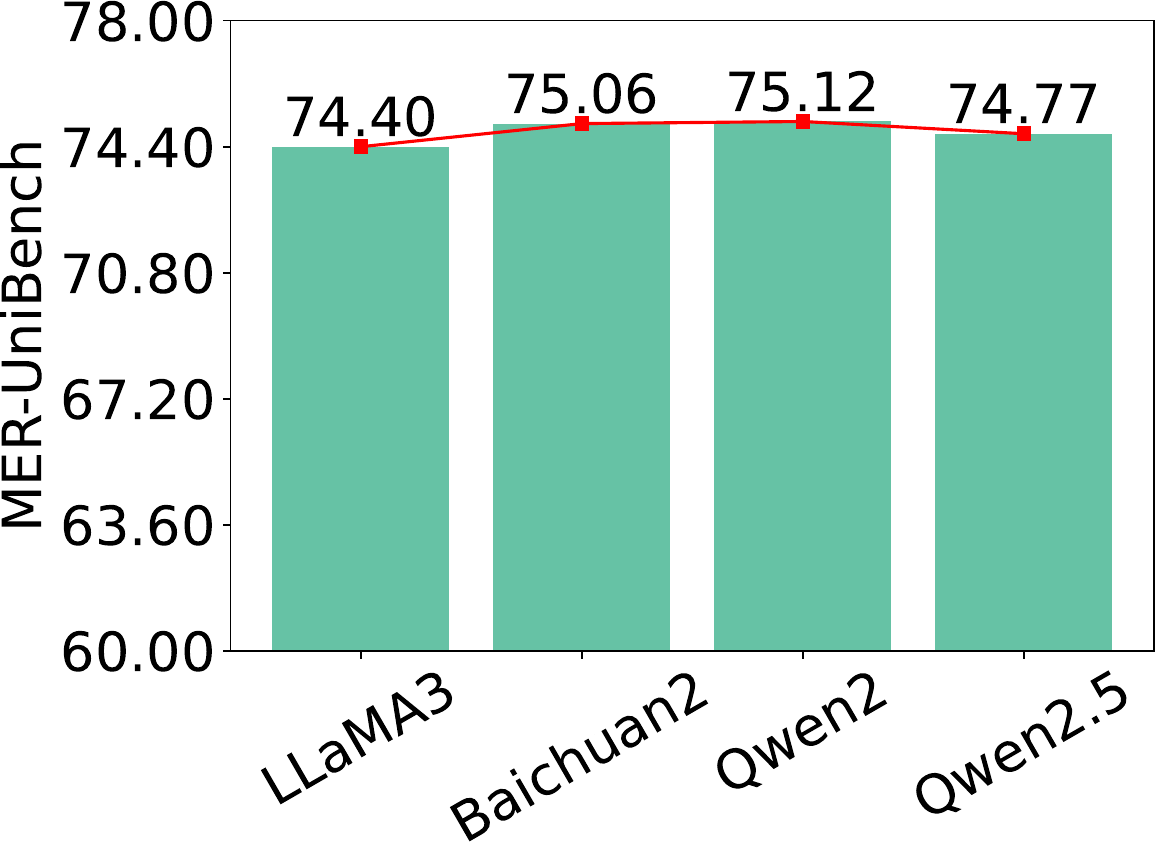}
		}
		\subfigure[\scriptsize{Audio Encoder}]{
			\label{Figure60-2}
			\centering
			\includegraphics[width=0.28\linewidth, trim=0 0 0 0]{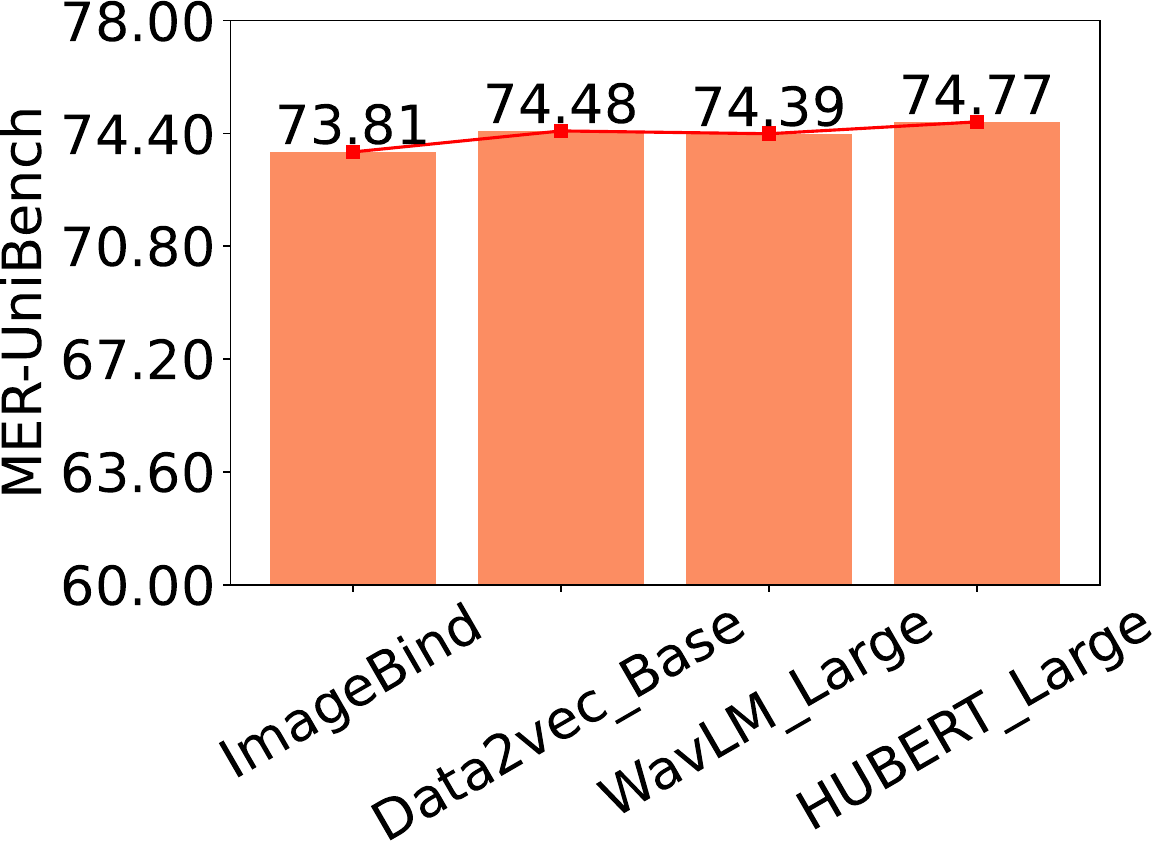}
		}
		\subfigure[\scriptsize{Video Encoder}]{
			\label{Figure60-3}
			\centering
			\includegraphics[width=0.28\linewidth, trim=0 0 0 0]{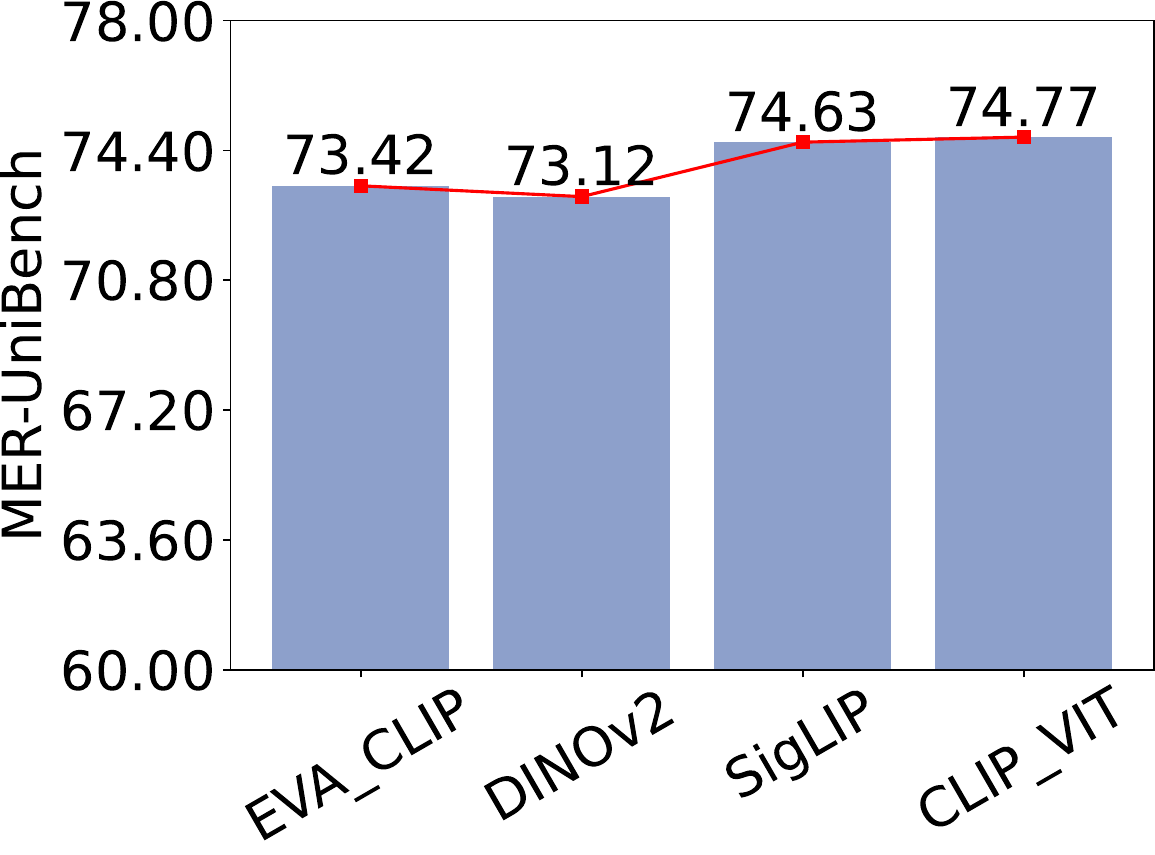}
		}
	\end{center}
	\caption{Ablation studies on LLMs, audio encoders, and video encoders.}
	\label{Figure60}
\end{figure}

\paragraph{Choice of LLMs.}
This paper adopts Qwen2.5 as the default LLM. In Figure \ref{Figure60-1}, we further explore the impact of different LLMs. Experimental results show that the performance difference brought by LLM is limited. These results verify that the superior performance of AffectGPT over the existing MLLMs does not come from LLM but from our proposed emotion description dataset and model.

\paragraph{Choice of Audio and Video Encoders.}
In Figures \ref{Figure60-2} and \ref{Figure60-3}, the choice of audio and video encoder has a minimal impact on performance. This underscores that AffectGPT's exceptional performance is primarily driven by our proposed high-quality, large-scale dataset and effective framework, rather than the specific acoustic or visual encoders used. For audio encoders (Figure \ref{Figure60-2}), ImageBind exhibits slightly inferior performance compared to other audio encoders. This may be attributed to the fact that other audio encoders are predominantly utilized in audio content understanding tasks (e.g., ASR), where audio content plays a critical role in emotion recognition. Similarly, for video encoders (Figure \ref{Figure60-3}), CLIP\_VIT marginally outperforms EVA\_CLIP and DINOv2, aligning with findings from MERBench \cite{lian2024merbench}, a unified benchmark for traditional categorical frameworks. These results suggest that insights derived from traditional categorical frameworks, such as encoder selection, may also be applicable to MLLM-based descriptive frameworks.

\paragraph{Role of LoRA in LLMs.}
In Table \ref{Table603}, we count the increase in trainable parameters when using LoRA for the LLM branch. The first row represents the model without the LoRA module. Experimental results show that fine-tuning the LLM with LoRA improves performance compared to models without LoRA. However, increasing the rank for LoRA-based models does not yield significant performance gains and instead increases computational costs. 

\begin{table}[t]
	\centering
	\caption{\textbf{Impact of rank value in LoRA.} In this table, we count the increase in trainable parameters when using LoRA for the LLM branch. The first row represents the model without the LoRA module.}
	\label{Table603}
	\scalebox{0.8}{
		\begin{tabular}{c|r|c}
			\hline
			\textbf{Rank}
			& \textbf{\# Increased Parameters} & \textbf{MER-UniBench} \\
			\hline	
                  --  & 0           & 73.30 \\
                  8   & 20,185,088  & 74.65 \\
                  16  & 40,370,176  & 74.77 \\
                  32  & 80,740,352  & 74.92 \\
			\hline
		\end{tabular}
	}
\end{table}

\section{Conclusion}
\label{sec:6}
This paper aims to enhance the emotional understanding of MLLMs from three aspects: (1) the dataset MER-Caption, which uses a model-led human-assisted strategy to create a large-scale dataset with guaranteed quality; (2) the model AffectGPT, which enhances multimodal fusion by moving cross-modal interactions outside of the LLM; and (3) the benchmark, which provides comprehensive evaluation metrics tailored to the free-form, natural language output style of MLLMs. Extensive experiments validate the effectiveness of our model and dataset. This work lays the foundation for building MLLMs with emotional understanding, contributing to the advancement of emotion AI.

\section*{Acknowledgments}
This work is supported by the Excellent Youth Program of State Key Laboratory of Multimodal Artificial Intelligence Systems (MAIS2024311), the National Natural Science Foundation of China (62201572, 62322120, 61831022, 62276259, U21B2010, 62271083, 62306316, 62176165, 62206136, 62476146), the Stable Support Projects for Shenzhen Higher Education Institutions (20220718110918001), the Young Elite Scientists Sponsorship Program by CAST (2024QNRC001), and the University of Oulu\& Research Council of Finland Profi 7 (352788).

\section*{Impact Statements}
\paragraph{Social Impact.}
Emotion plays an important role in human communication, conveying human intentions and deep thoughts. As Minsky \cite{minsky1988society} stated: \emph{The question is not whether intelligent machines can have any emotions, but whether machines can be intelligent without any emotions.} The development of MER technology can enhance the human-computer interaction experience.

\paragraph{Ethics Statement.}
This paper does not involve the collection of new data. The original data comes from the unlabeled part of MER2024 \cite{lian2024mer}, with permission from the dataset owners. The annotation process does not involve hiring external annotators, and no ethical issues are associated with this process. Additionally, we restrict the use of this dataset under the license of CC BY-NC 4.0, requiring researchers to use our dataset responsibly. Therefore, no ethical concerns are raised in this paper.

\bibliography{mybib}
\bibliographystyle{icml2025}

\newpage
\appendix
\onecolumn
\section{Related Works}
This paper focuses on constructing datasets and designing models to enhance the emotional understanding capability of MLLMs. In this section, we mainly review related work in these two aspects.

\subsection{Emotion Dataset}
Emotion datasets are the foundation for building MER systems \cite{wang2022ferv39k, chen2023smg}. Most research has focused on building categorical datasets, where basic emotions are first defined, and annotators are asked to select the most likely one \cite{goodfellow2013challenges} or multiple \cite{li2017reliable} labels from basic emotions. However, emotions are often diverse \cite{demszky2020goemotions} and can coexist \cite{du2014compound}, making it challenging for categorical datasets to fully capture these complex emotions.

To address this, recent studies have shifted from categorical datasets to descriptive datasets, as emotion descriptions provide greater flexibility and enable the description of complex emotions in natural language. To construct such datasets, \citet{liu2022mafw} used a human-based annotation strategy to capture the environment, body movements, facial expressions, and other emotion-related cues. However, the high annotation cost limits the scalability of these datasets. With the development of MLLMs, \citet{cheng2024emotion} used a more cost-effective automatic annotation method, where MLLMs are used to extract emotion-related descriptions from audio, facial expressions, and visual objects. However, they lacked pre-experimentation on MLLM selection, relying on empirical model choices, leading to insufficient label quality. In this paper, we propose a solution to balance label quality and dataset size. By leveraging high-quality human-based datasets to guide description generation and sample filtering, we achieve a quality-assured automatic annotation process and ultimately construct MER-Caption.

\subsection{Emotion Models}
Emotion models are closely related to the training corpus. For categorical datasets, researchers often build classifiers to map multimodal human information to corresponding emotion labels. Apart from choosing the architecture (such as CNN, RNN, or Transformers), most research focuses on how to align and fuse multimodal information. For example, \citet{hazarika2020misa} introduced a decomposition module to split features into modality-specific and modality-invariant representations. \citet{gu2018multimodal} aligned different modalities at the word level and then learned time-dependent cross-modal interactions. \citet{tsai2019multimodal} proposed using cross-attention to align features in the latent space. More recently, \citet{lian2024merbench} conducted a fair comparison of various fusion and alignment strategies, showing that temporal-preserving features do not always outperform time-compressed features, suggesting that MER may be more suitable to solve from a global perspective.

For descriptive datasets, due to their natural language style output, the framework needs to shift from traditional discriminative methods to generative methods. With the development of LLMs and MLLMs, researchers have started to build models based on them. For example, \citet{huang2024ecr} used Vicuna as the language model, jointly training emotion labels and descriptions. \citet{xie2024emovit} used the instruction-aware Q-Former module to learn the mapping between input images and emotional descriptions. \citet{cheng2024emotion} integrated different encoders to understand multimodal inputs and used LLaMA-2 as an LLM decoder. However, current models either only focus on unimodal information \cite{huang2024ecr, xie2024emovit} or leave all cross-modal interactions to the LLM \cite{cheng2024emotion}, which is insufficient for solving MER tasks with multimodal characteristics. To this end, we introduce the AffectGPT model in this paper.

\section{Implementation Details}
\label{appendix:implementation_details}
Our choice of unimodal encoders is guided by previous research \cite{lian2024merbench}, using CLIP ViT-L \cite{radford2021learning} as the visual encoder and HUBERT-L \cite{hsu2021hubert} as the acoustic encoder. Given the remarkable performance of Qwen-2.5 \cite{yang2024qwen2}, we choose it as the LLM. To ensure training efficiency, we only fine-tune an extra LoRA module (in the LLM), projector, and pre-fusion branch, while freezing the weights of the LLM and unimodal encoders (see Figure \ref{Figure3}). We default to setting the rank in the LoRA module to 16. This approach reduces GPU memory usage and speeds up training. Additionally, through preliminary experiments, we found that pre-training on other instruction datasets followed by a second-stage training on MER-Caption did not lead to performance improvements. The primary reason is the large scale of our dataset and the limited focus on MER in current instruction datasets. Therefore, we do not perform multi-stage training in our experiments. All models are implemented in PyTorch and conducted training and inference on 80GB NVIDIA Tesla A100 GPU. During training, we set the maximum number of epochs to 60, each epoch contains 5000 iterations, and the batch size of each iteration is 3. To optimize all trainable parameters, we use the AdamW optimizer and set the learning rate to 1e-5. For more implementation details, please refer to the code provided in \href{https://github.com/zeroQiaoba/AffectGPT}{https://github.com/zeroQiaoba/AffectGPT}.

\section{Details about MLLMs}
\label{appendix_sec:mllm}
Table \ref{Table12} provides model cards for different MLLMs, including reference papers, supported modalities, and links to pre-trained weights.

\begin{table}[h]
	\centering
	\caption{Model cards for MLLMs.}
	\label{Table12}
	\scalebox{0.8}{
		\begin{tabular}{l|l|l}
			\hline
			 & \textbf{Supported Modality} & \textbf{Link} \\
			\hline
			Otter \cite{li2023otter} & Video, Text & \textcolor[rgb]{0.93,0.0,0.47}{https://github.com/Luodian/Otter} \\
			VideoChat \cite{li2023videochat} & Video, Text & \textcolor[rgb]{0.93,0.0,0.47}{https://github.com/OpenGVLab/Ask-Anything/tree/main/video\_chat} \\
			VideoChat2 \cite{li2024mvbench} & Video, Text & \textcolor[rgb]{0.93,0.0,0.47}{https://github.com/OpenGVLab/Ask-Anything/tree/main/video\_chat2} \\
			Video-LLaVA \cite{lin2024video} & Video, Text & \textcolor[rgb]{0.93,0.0,0.47}{https://github.com/PKU-YuanGroup/Video-LLaVA} \\
			Video-LLaMA \cite{zhang2023video} & Video, Text & \textcolor[rgb]{0.93,0.0,0.47}{https://github.com/DAMO-NLP-SG/Video-LLaMA} \\
			Video-ChatGPT \cite{maaz2024video} & Video, Text & \textcolor[rgb]{0.93,0.0,0.47}{https://github.com/mbzuai-oryx/Video-ChatGPT} \\
			LLaMA-VID \cite{li2024llama} & Video, Text & \textcolor[rgb]{0.93,0.0,0.47}{https://github.com/dvlab-research/LLaMA-VID} \\
			mPLUG-Owl \cite{ye2023mplug} & Video, Text & \textcolor[rgb]{0.93,0.0,0.47}{https://github.com/X-PLUG/mPLUG-Owl} \\
			Chat-UniVi \cite{jin2024chat} & Video, Text & \textcolor[rgb]{0.93,0.0,0.47}{https://github.com/PKU-YuanGroup/Chat-UniVi} \\
			SALMONN \cite{tang2023salmonn} & Audio, Text & \textcolor[rgb]{0.93,0.0,0.47}{https://github.com/bytedance/SALMONN} \\
			Qwen-Audio \cite{chu2023qwen} & Audio, Text & \textcolor[rgb]{0.93,0.0,0.47}{https://github.com/QwenLM/Qwen-Audio} \\
			SECap \cite{xu2024secap} & Audio, Text & \textcolor[rgb]{0.93,0.0,0.47}{https://github.com/thuhcsi/SECap} \\
			OneLLM \cite{han2024onellm} & Audio, Video, Text & \textcolor[rgb]{0.93,0.0,0.47}{https://github.com/csuhan/OneLLM} \\
			PandaGPT \cite{su2023pandagpt} & Audio, Video, Text & \textcolor[rgb]{0.93,0.0,0.47}{https://github.com/yxuansu/PandaGPT} \\
                Emotion-LLaMA \cite{cheng2024emotion} & Audio, Video, Text & \textcolor[rgb]{0.93,0.0,0.47}{https://github.com/ZebangCheng/Emotion-LLaMA} \\
			\hline
			
		\end{tabular}
	}
\end{table}

\clearpage
\section{Visualization of MLLM Outputs}
\label{appendix:free_form}
\begin{figure*}[h]
	\centering
	\includegraphics[width=\linewidth]{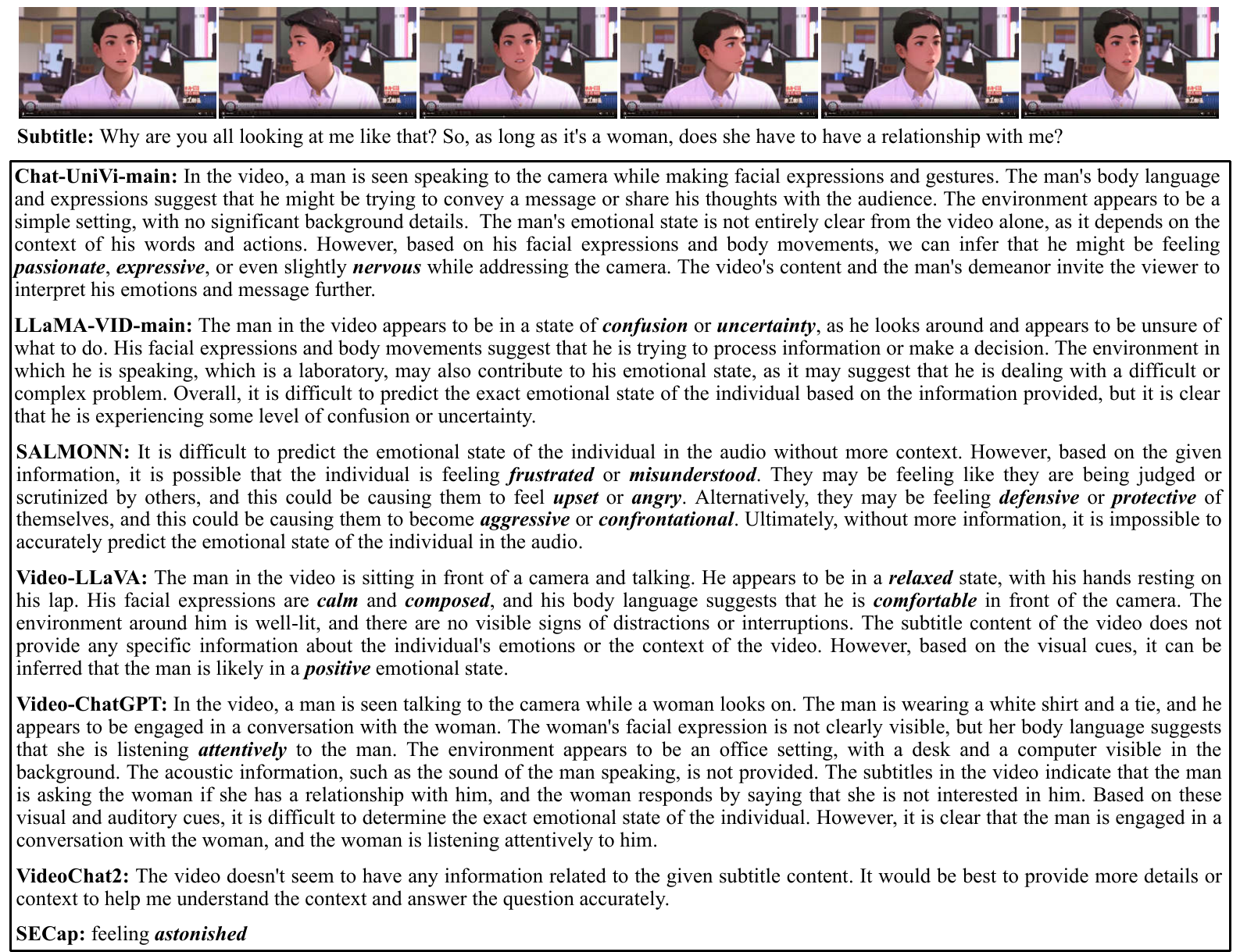}
	\caption{Visualization of MLLM outputs.}
	\label{Figure5}
\end{figure*}

Figure \ref{Figure5} provides an example to visualize the outputs of different MLLMs. These outputs contain varying numbers of emotions, with emotion labels that are open-ended and not restricted to any predefined taxonomy. Therefore, traditional classification metrics, such as accuracy and F1 score, are not suitable for performance evaluation. In this paper, we propose evaluation metrics specifically tailored for the free-form, natural language output style of MLLMs.

\section{Prompt for Label Extraction}
\label{appendix:label_extraction_prompt}
To extract emotion labels from MLLM outputs, we use Qwen2.5 and apply the following prompt:

\textcolor[rgb]{0.93,0.0,0.47}{\emph{Please assume the role of an expert in the field of emotions. We provide clues that may be related to the emotions of the characters. Based on the provided clues, please identify the emotional states of the main character. Please separate different emotional categories with commas and output only the clearly identifiable emotional categories in a list format. If none are identified, please output an empty list.}}

For sentiment analysis, we use the multi-step prediction process. Specifically, we first extract emotion labels using the prompt above, and then apply the following prompt for sentiment analysis:

\textcolor[rgb]{0.93,0.0,0.47}{\emph{Please act as an expert in the field of emotions. We provide a few words to describe the emotions of a character. Please choose the most likely sentiment from the given candidates: [positive, negative, neutral].}}

\section{Choice of Description Generation Strategy}
\label{appendix:multimodal_fusion}
This section aims to determine the optimal strategy for generating descriptions. In Table \ref{Table200}, we present the results of preliminary experiments. First, we evaluate the performance of different ALLMs and VLLMs. Then, we investigate whether combining these models leads to improved performance. To do this, we use GPT-3.5 to integrate audio and video cues, extracted by the ALLM and VLLM, with text content. As shown in Table \ref{Table200}, we observe that these combinations generally outperform the use of either ALLM or VLLM alone. Based on these findings, we select SALMONN as the ALLM for generating audio cues, Chat-UniVi as the VLLM for generating visual cues, and GPT-3.5 to combine the audio, video, and text cues, resulting in the final descriptions. 

We would like to clarify that, in this paper, we do not use \emph{combined results} for model selection. Instead, we rely on the performance of \emph{individual models}. Specifically, for VLLM, Chat-UniVi outperforms mPLUG-Owl and Video-ChatGPT; for ALLM, SALMONN outperforms SECap. As a result, we employ the combination of Chat-UniVi and SALMONN for description generation. The combination experiments are primarily designed to demonstrate that integrating multimodal cues can enhance performance. In future work, we will conduct additional experiments where \emph{combined results} are used for model selection. For example, leveraging the combination of SALMONN and Chat-UniVi for description generation.

\begin{table*}[h]
	\centering
	\renewcommand\tabcolsep{10pt}
	\caption{\textbf{Preliminary experiments.} We choose $\mbox{F}_{\mbox{s}}$ as the primary metric, as this metric considers both accuracy and completeness.}
	\label{Table200}
	\scalebox{0.9}{
		\begin{tabular}{l|>{\columncolor{lightgray}}ccc}
			\hline
			{{Model}} &$\mbox{F}_{\mbox{s}}$($\uparrow$) & $\mbox{Precision}_{\mbox{s}}$($\uparrow$) & $\mbox{Recall}_{\mbox{s}}$($\uparrow$) \\
            \hline
			SECap \cite{xu2024secap}          &45.72$_{\pm0.09}$ & 54.52$_{\pm0.15}$ & 39.37$_{\pm0.05}$ 	 \\
SALMONN  \cite{tang2023salmonn}       &47.96$_{\pm0.04}$ & 50.20$_{\pm0.04}$ & 45.92$_{\pm0.04}$ 	 \\
Video-ChatGPT \cite{maaz2024video}  &50.52$_{\pm0.06}$ & 54.03$_{\pm0.04}$ & 47.44$_{\pm0.07}$ 	 \\
mPLUG-Owl \cite{ye2023mplug}     &52.73$_{\pm0.13}$ & 54.54$_{\pm0.13}$ & 51.04$_{\pm0.13}$ 	 \\
Chat-UniVi \cite{jin2024chat}     &53.08$_{\pm0.01}$ & 53.68$_{\pm0.00}$ & 52.50$_{\pm0.02}$ 	 \\
\hline
SECap + mPLUG-Owl       &56.69$_{\pm0.03}$ & 50.05$_{\pm0.23}$ & 65.38$_{\pm0.33}$ \\
SECap + Video-ChatGPT   &56.90$_{\pm0.08}$ & 52.03$_{\pm0.04}$ & 62.79$_{\pm0.14}$ \\
SECap + Chat-UniVi      &57.34$_{\pm0.16}$ & 48.85$_{\pm0.29}$ & 69.41$_{\pm0.13}$ \\
SALMONN + Video-ChatGPT &58.19$_{\pm0.23}$ & 53.16$_{\pm0.17}$ & 64.26$_{\pm0.31}$ \\
SALMONN + Chat-UniVi    &58.43$_{\pm0.06}$ & 51.62$_{\pm0.00}$ & 67.31$_{\pm0.15}$ \\
SALMONN + mPLUG-Owl     &\textbf{58.70}$_{\pm0.04}$ & 51.77$_{\pm0.01}$ & 67.76$_{\pm0.11}$ \\
\hline
		\end{tabular}
	}
\end{table*}

\clearpage
\section{Prompt for Clue Merge}
\label{appendix:clue_merge_prompt}
To merge multimodal clues, we use GPT-3.5 and apply the following prompt:

\textcolor[rgb]{0.93,0.0,0.47}{\emph{Please act as an expert in the field of emotions. We provide acoustic and visual clues that may be related to the character's emotional state, along with the original subtitle of the video. Please analyze which parts can infer the emotional state and explain the reasons. During the analysis, please integrate the textual, audio, and visual clues.}}

\begin{figure*}[h]
	\centering
	\includegraphics[width=\linewidth]{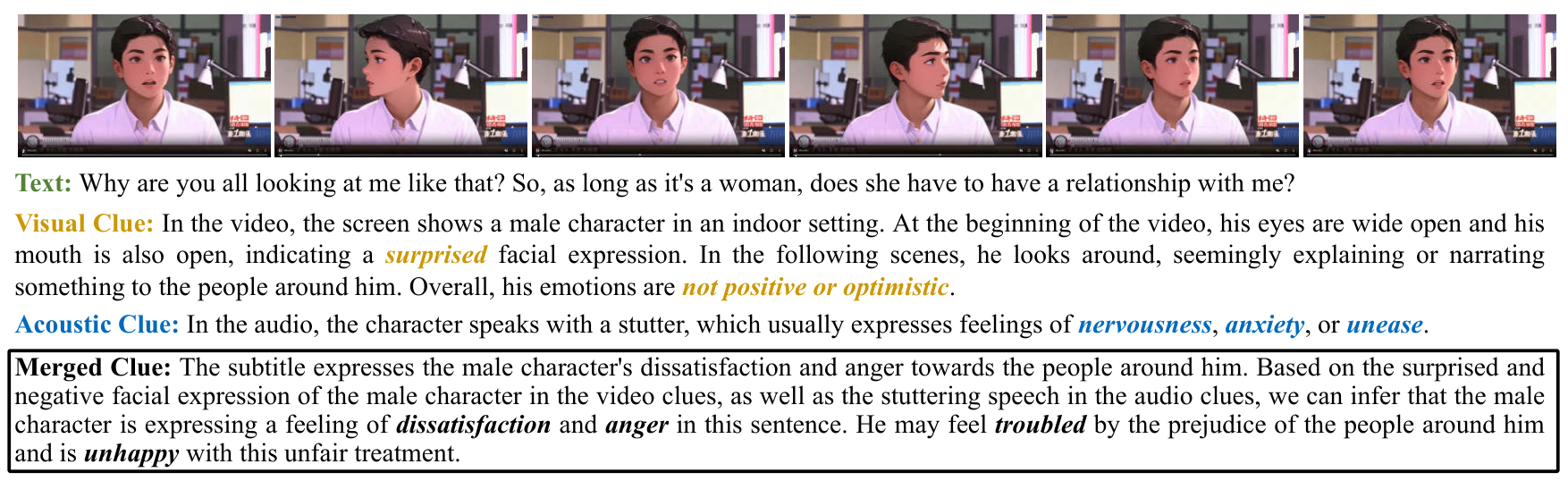}
	\caption{Example of modality conflict.}
	\label{Figure4}
\end{figure*}

Even when modality conflicts exist (i.e., the emotions conveyed by audio, video, and text are not the same, as shown in Figure \ref{Figure4}), GPT-3.5 can provide reasonable responses, primarily due to its powerful reasoning ability.

\section{Dataset Comparison}
\label{appendix:dataset_comparison}
\begin{figure*}[h]
	\begin{center}
		\subfigure[{EmoVIT}]{
			\label{Figure8-1}
			\centering
			\includegraphics[width=0.139\linewidth, trim=30 0 30 0]{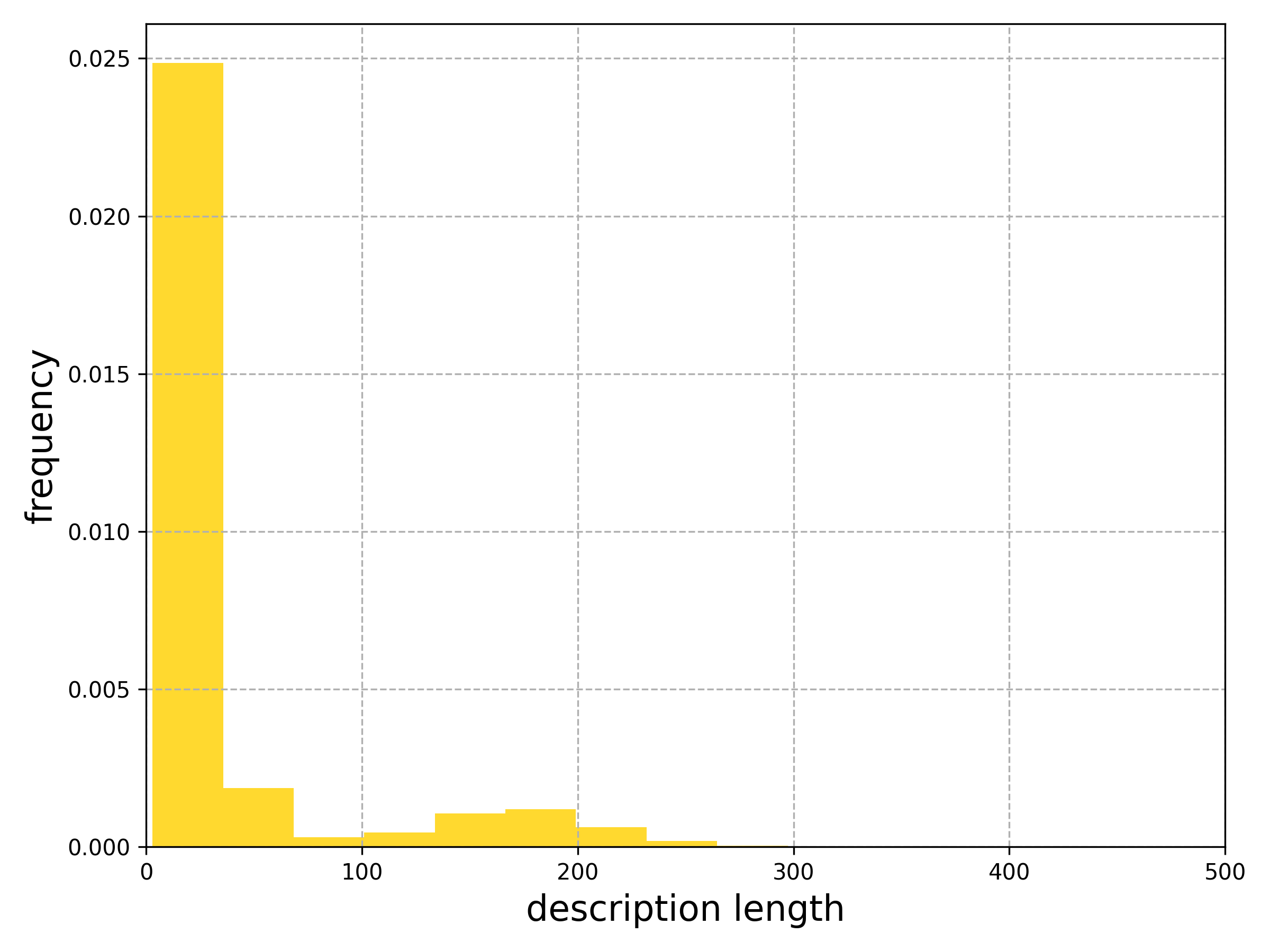}
		}
		\subfigure[{MERR-Fine}]{
			\label{Figure8-2}
			\centering
			\includegraphics[width=0.139\linewidth, trim=30 0 30 0]{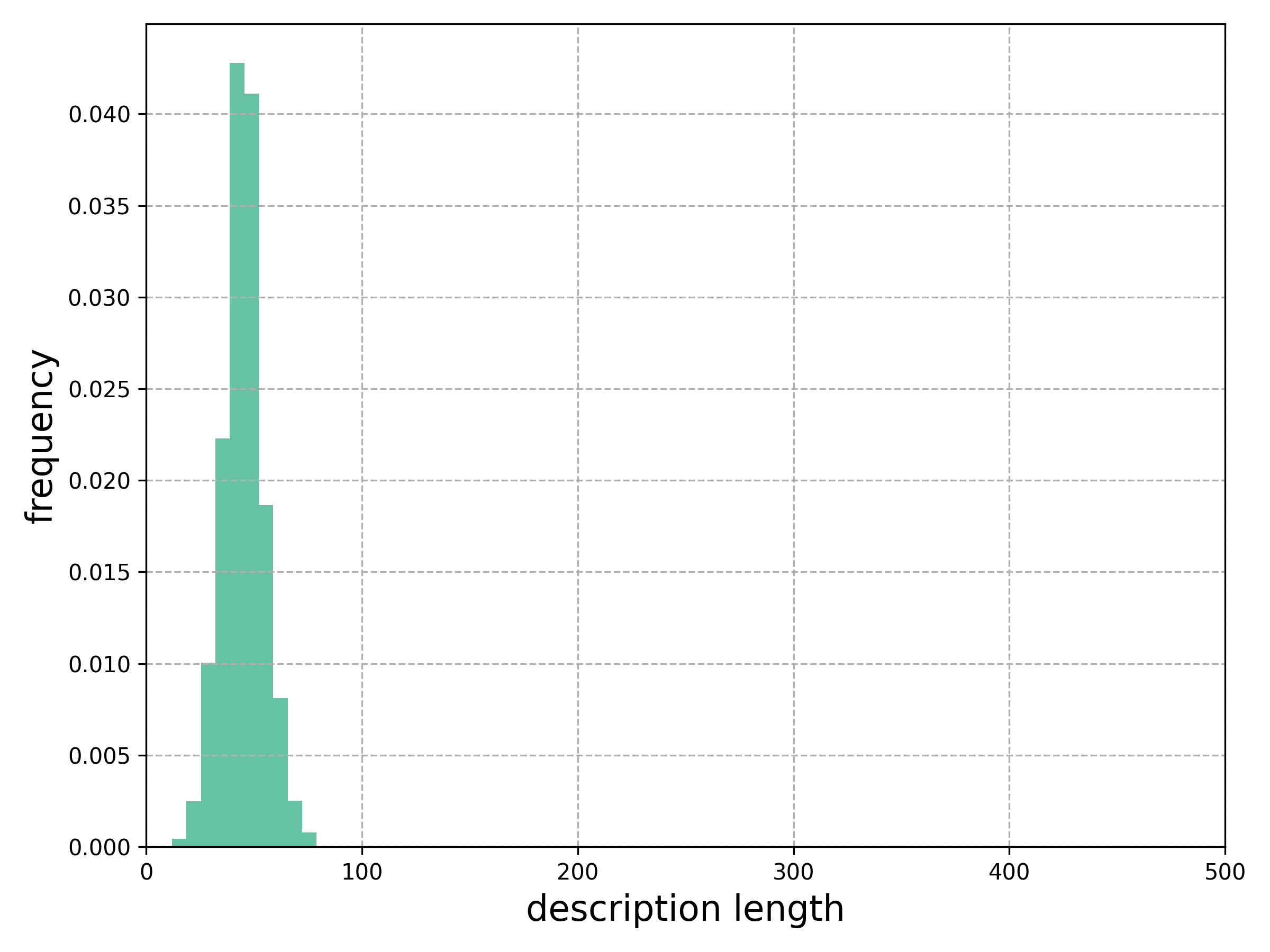}
		}
		\subfigure[{MERR-Coarse}]{
			\label{Figure8-3}
			\centering
			\includegraphics[width=0.139\linewidth, trim=30 0 30 0]{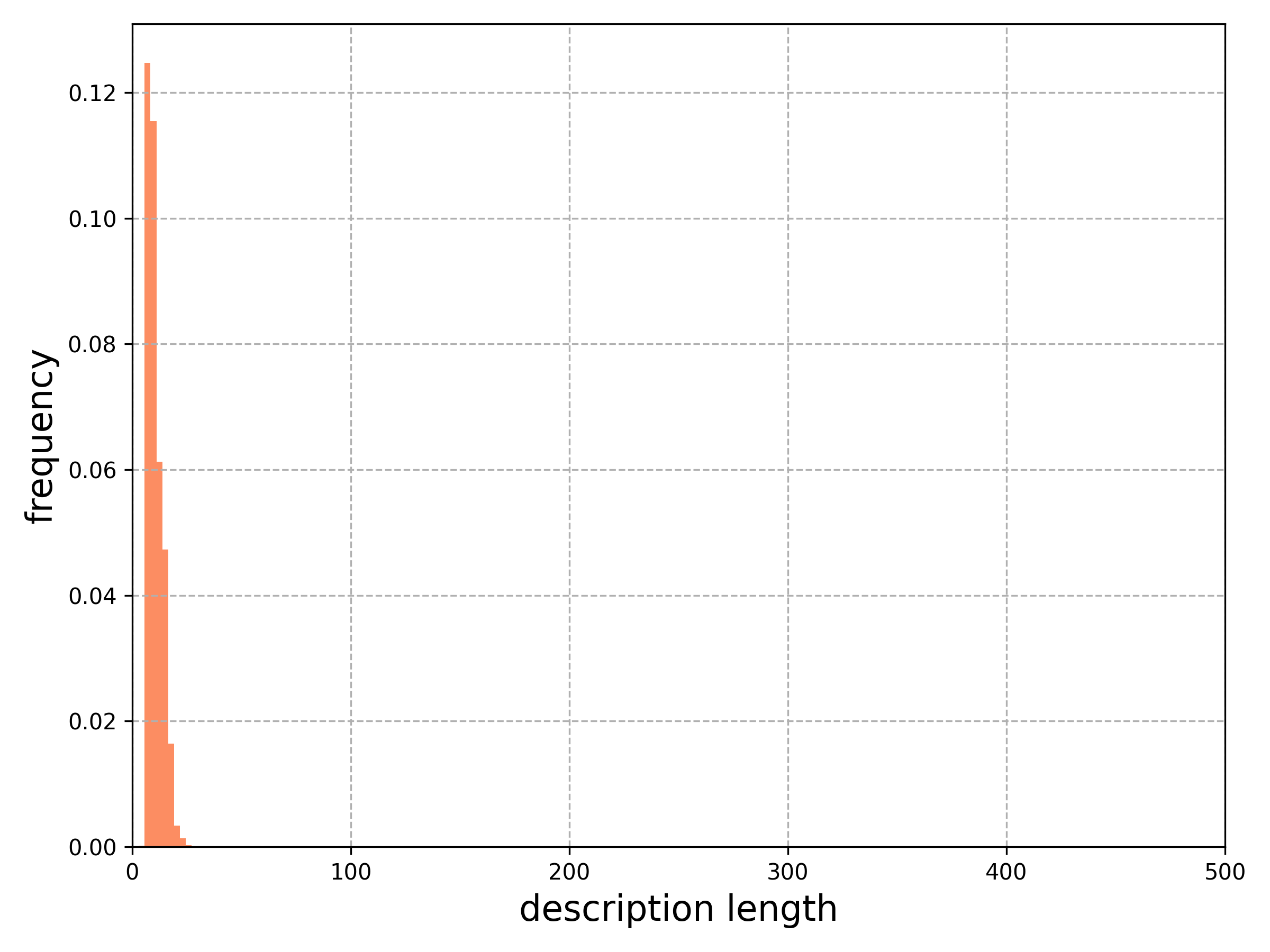}
		}
		\subfigure[{MAFW}]{
			\label{Figure8-4}
			\centering
			\includegraphics[width=0.139\linewidth, trim=30 0 30 0]{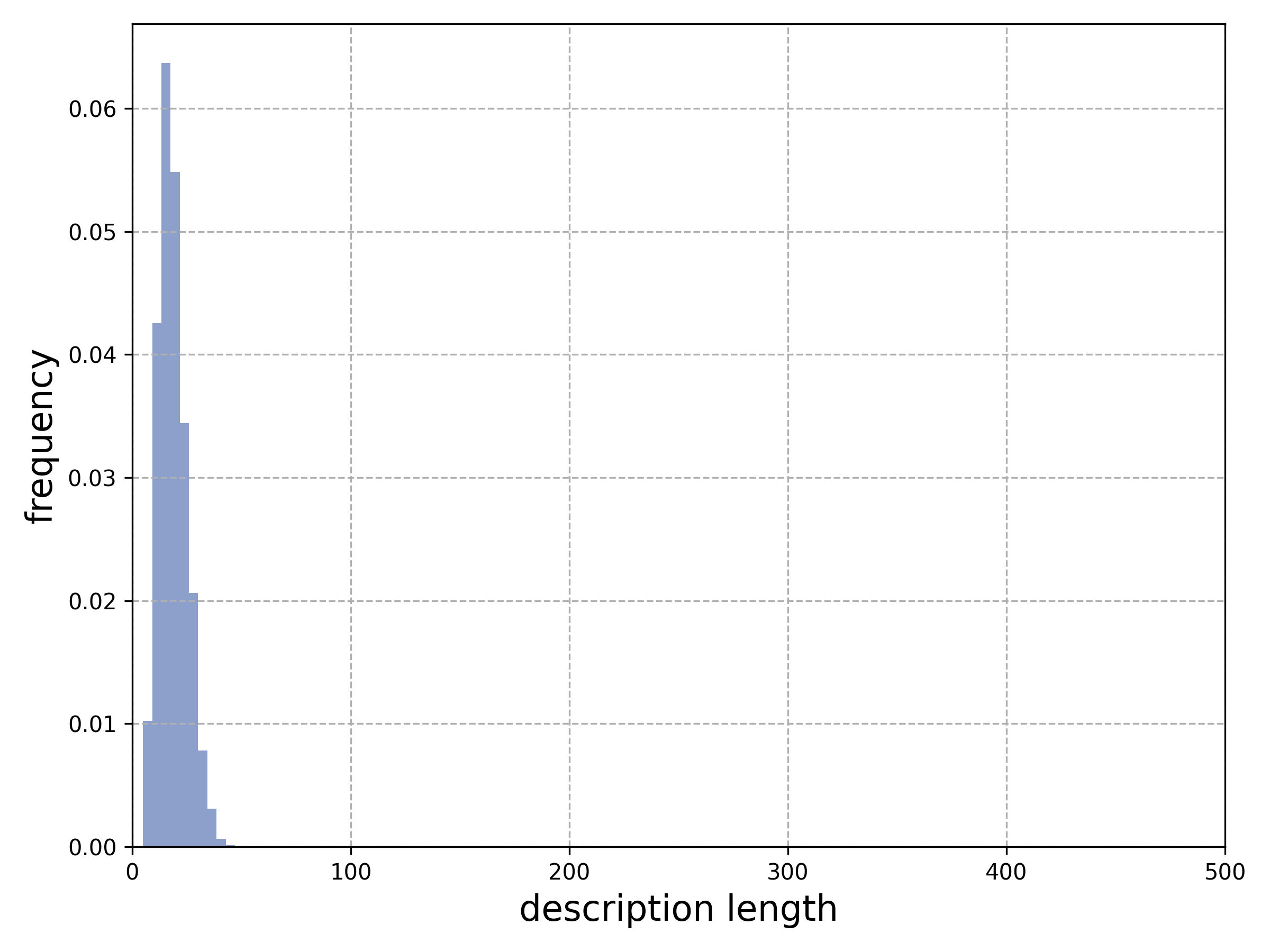}
		}
		\subfigure[{OV-MERD}]{
			\label{Figure8-5}
			\centering
			\includegraphics[width=0.139\linewidth, trim=30 0 30 0]{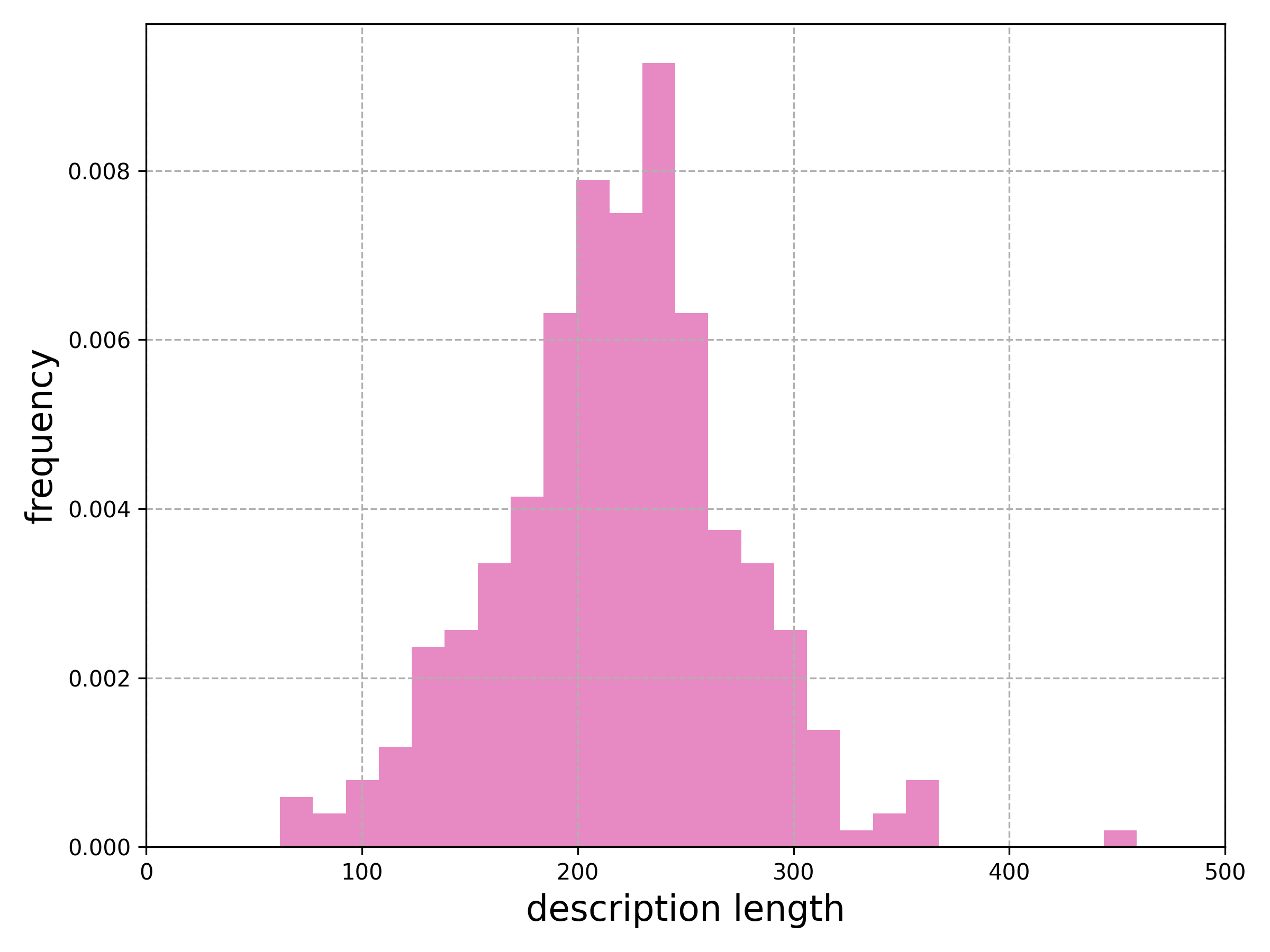}
		}
		\subfigure[{MER-Caption}]{
			\label{Figure8-6}
			\centering
			\includegraphics[width=0.139\linewidth, trim=30 0 30 0]{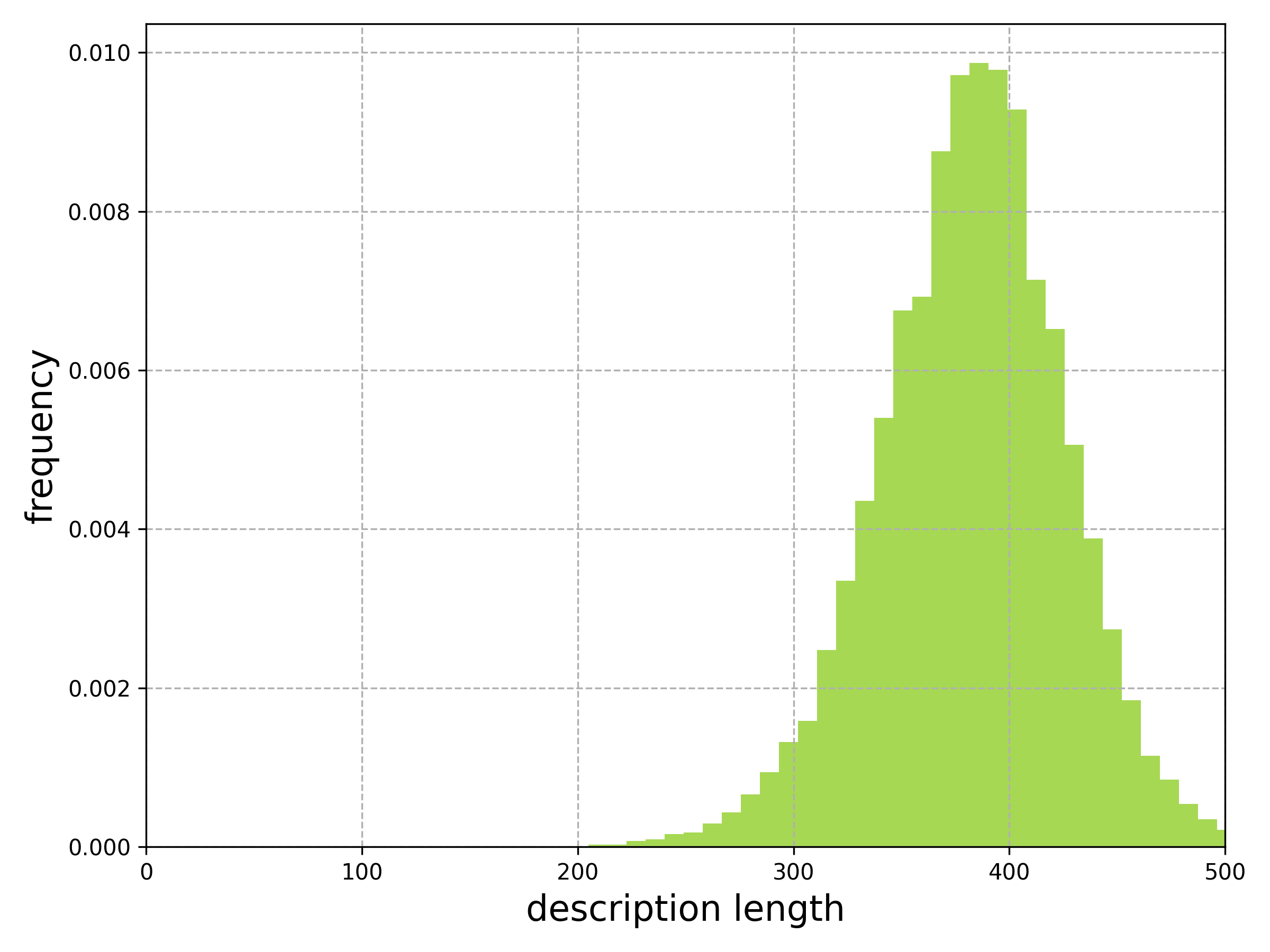}
		}
		
		\subfigure[{EmoVIT}]{
			\label{Figure8-7}
			\centering
			\includegraphics[width=0.139\linewidth, trim=30 0 30 0]{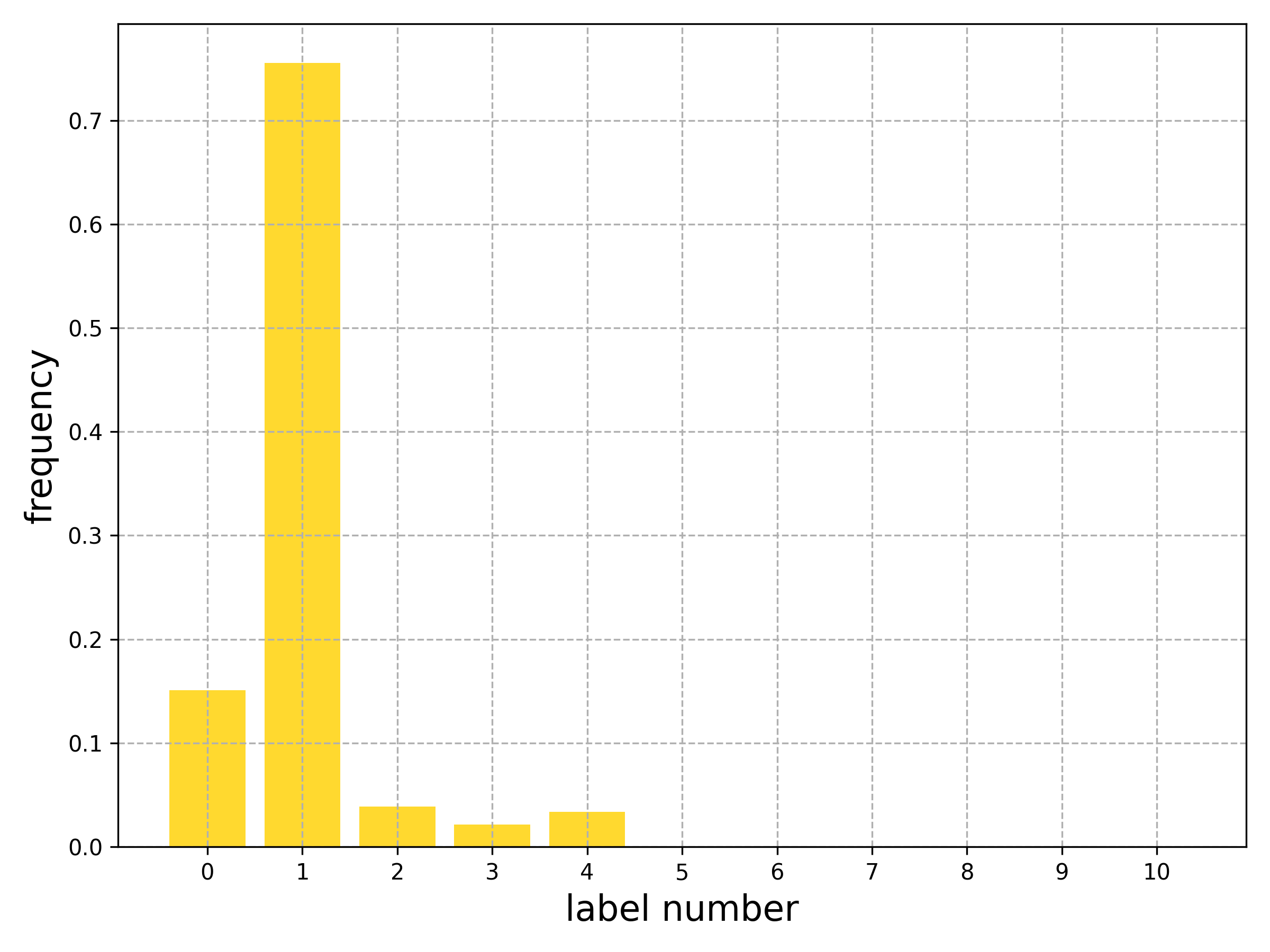}
		}
		\subfigure[{MERR-Fine}]{
			\label{Figure8-8}
			\centering
			\includegraphics[width=0.139\linewidth, trim=30 0 30 0]{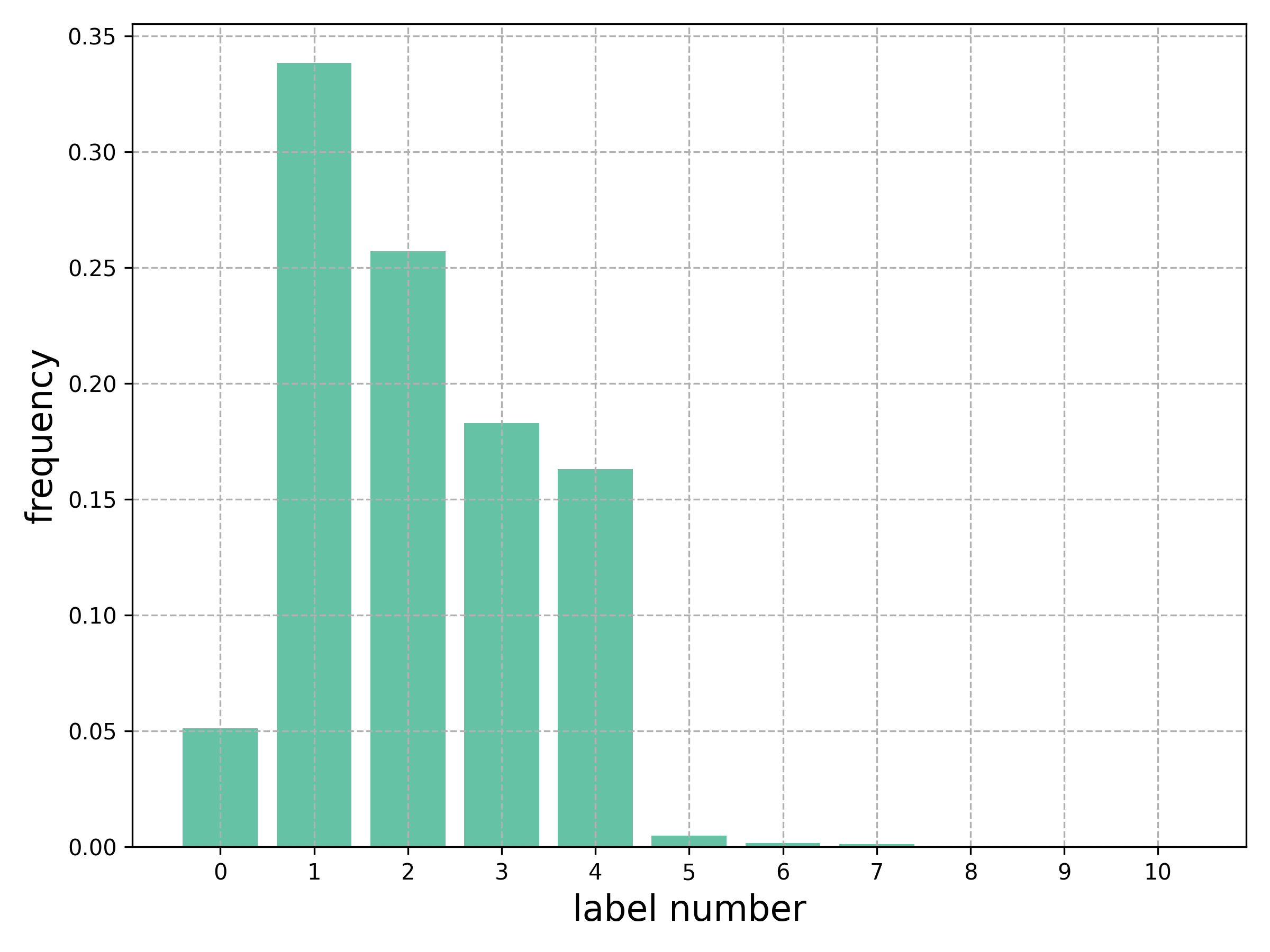}
		}
		\subfigure[{MERR-Coarse}]{
			\label{Figure8-9}
			\centering
			\includegraphics[width=0.139\linewidth, trim=30 0 30 0]{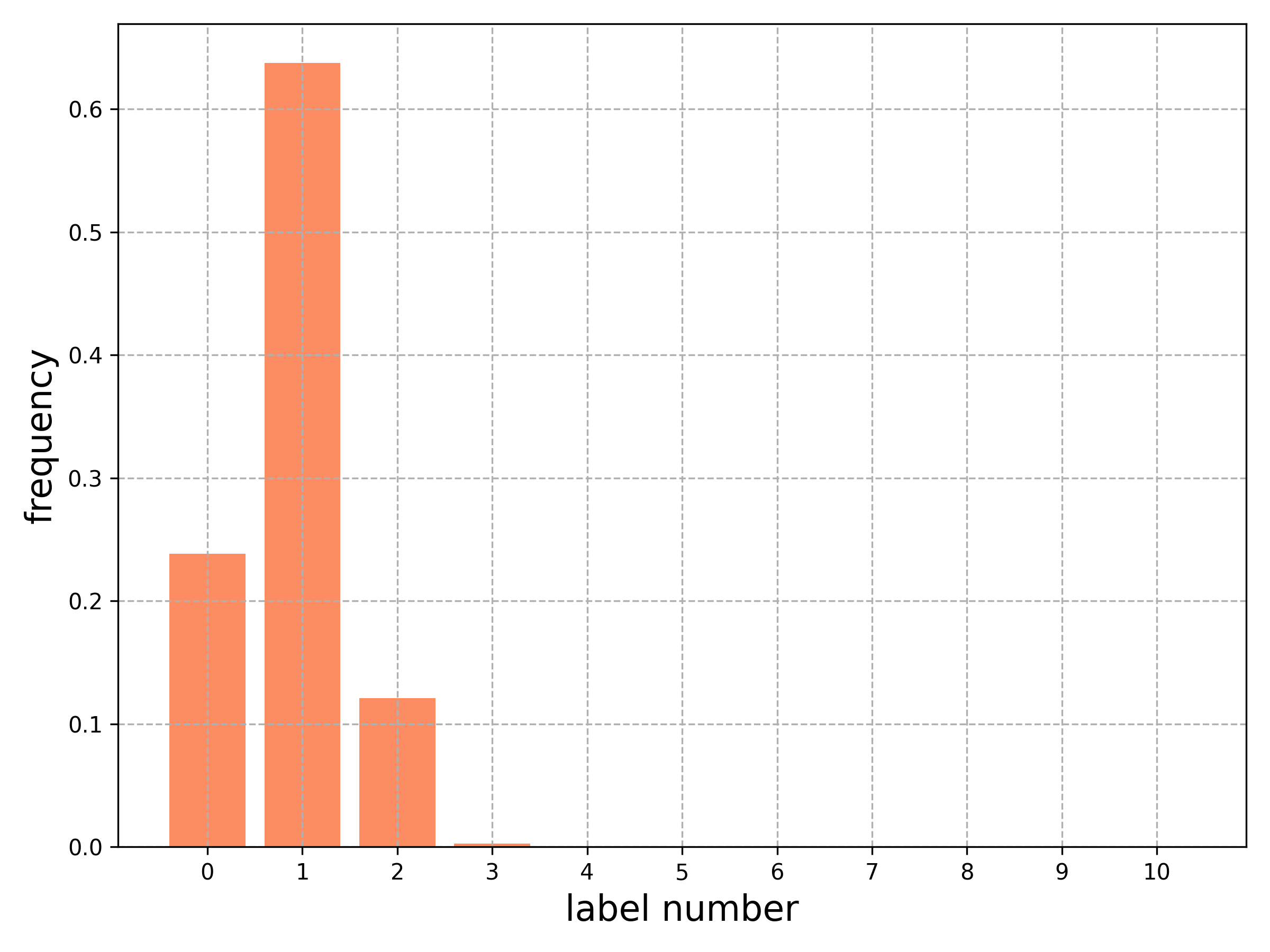}
		}
		\subfigure[{MAFW}]{
			\label{Figure8-10}
			\centering
			\includegraphics[width=0.139\linewidth, trim=30 0 30 0]{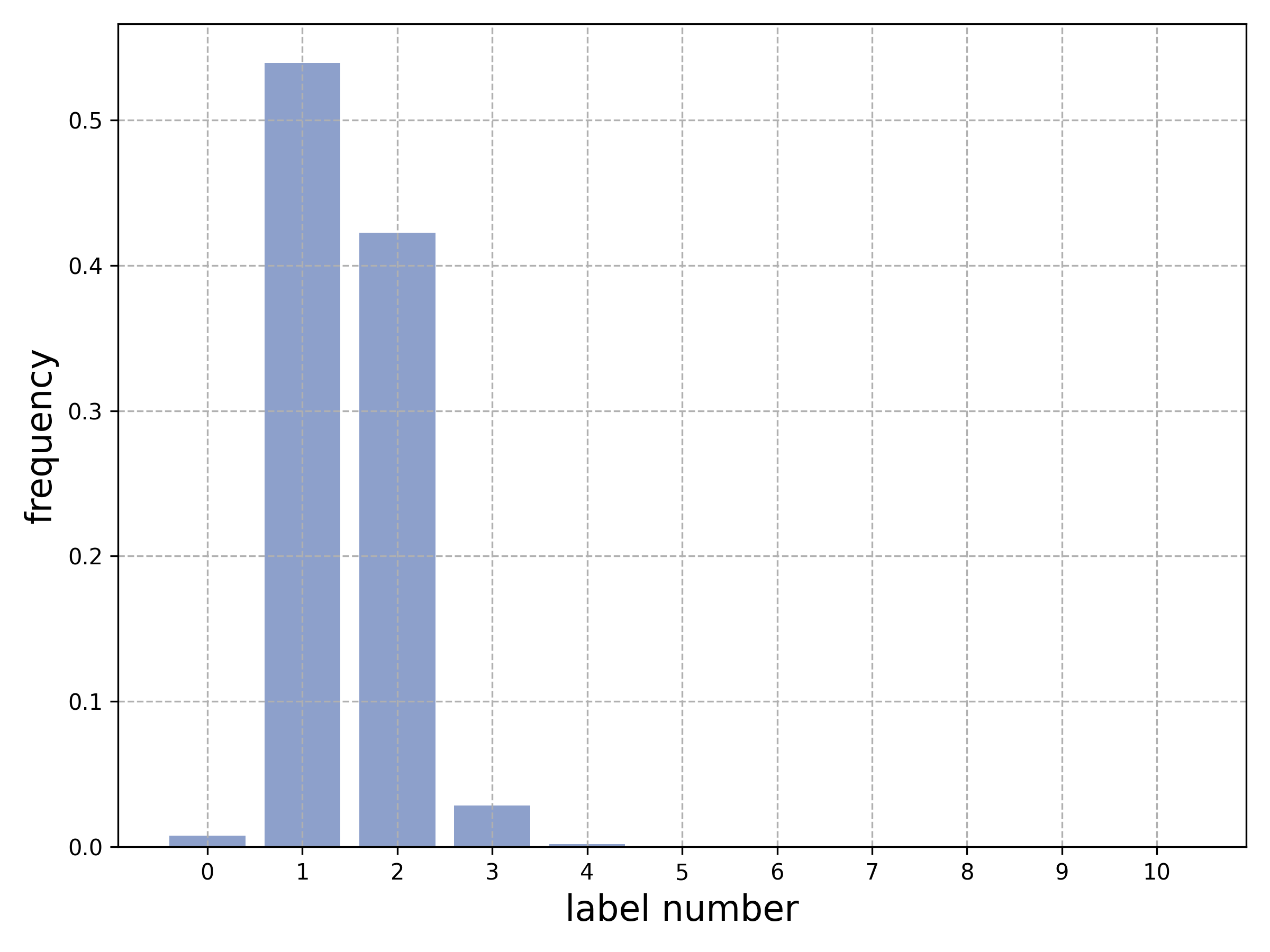}
		}
		\subfigure[{OV-MERD}]{
			\label{Figure8-11}
			\centering
			\includegraphics[width=0.139\linewidth, trim=30 0 30 0]{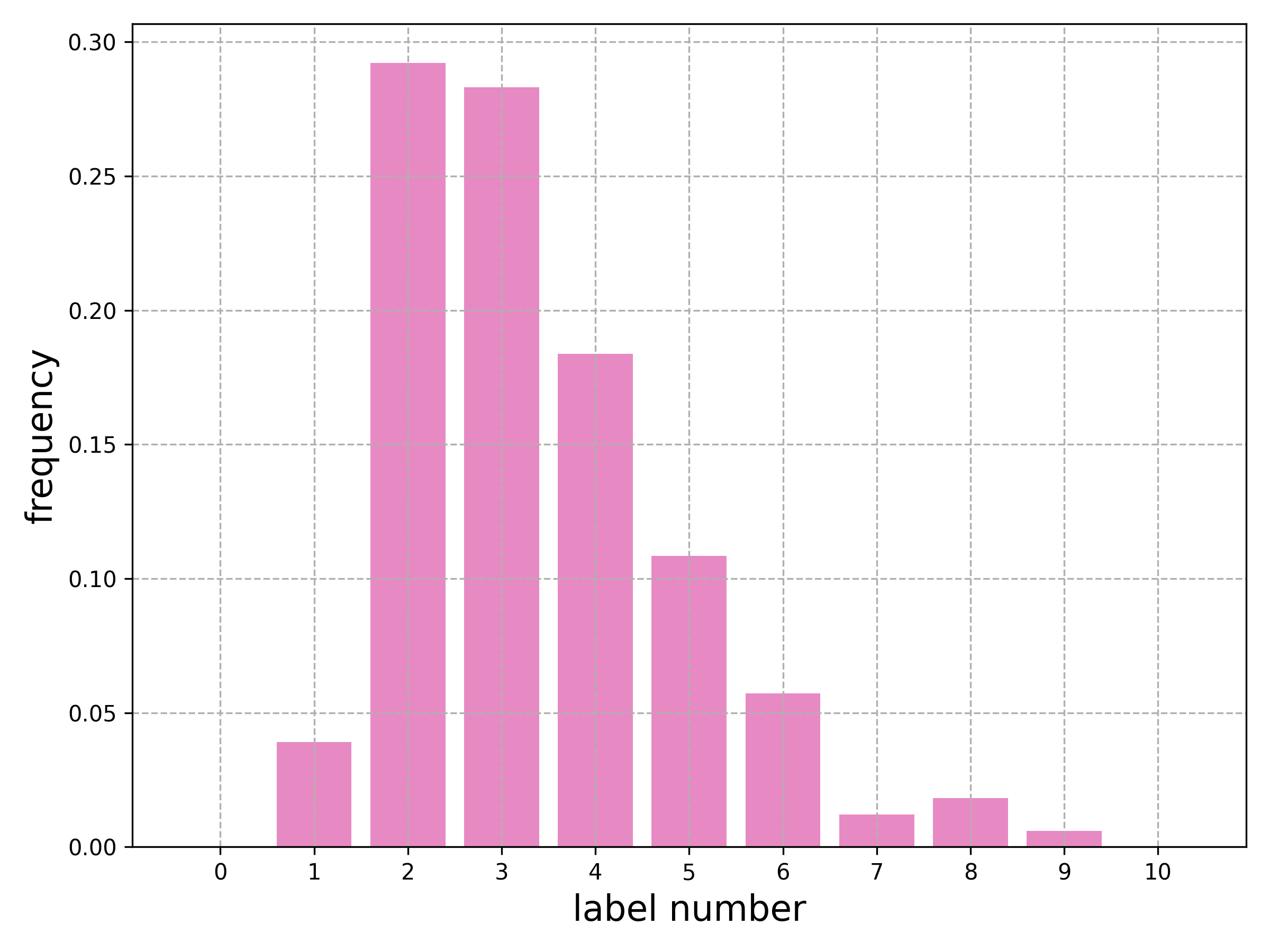}
		}
		\subfigure[{MER-Caption}]{
			\label{Figure8-12}
			\centering
			\includegraphics[width=0.139\linewidth, trim=30 0 30 0]{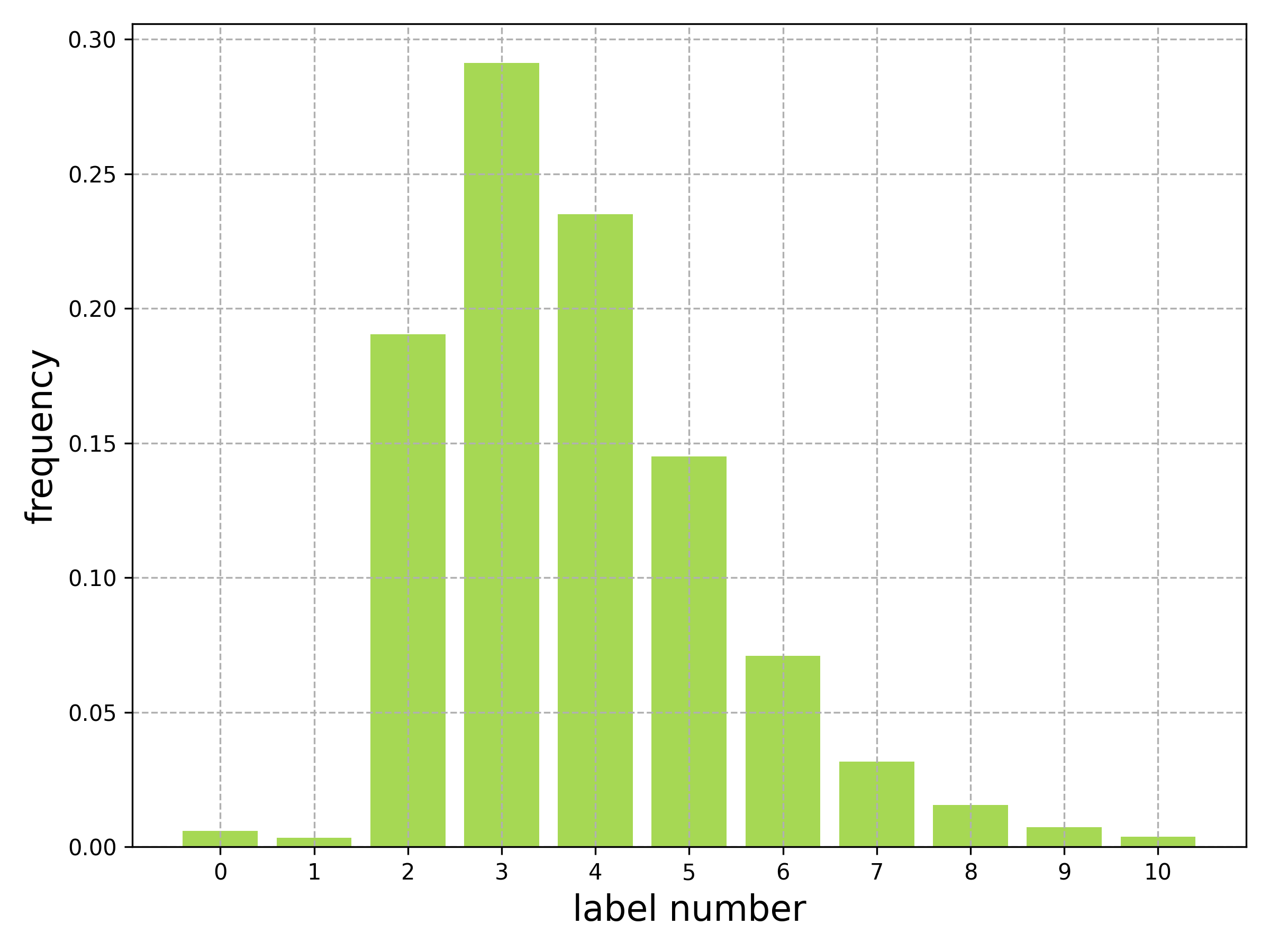}
		}
	\end{center}
	\caption{\textbf{Dataset comparison}. The first row compares the lengths of the descriptions, while the second row compares the number of labels per sample.}
	\label{Figure8}
\end{figure*}

Figure \ref{Figure8} compares the distribution of description lengths and the number of emotions per sample. We observe that our dataset provides detailed descriptions and rich emotion labels for each sample.

\section{Video Duration Distribution}
\label{appendix:video_duration}
\begin{figure*}[h]
	\centering
	\includegraphics[width=0.36\linewidth]{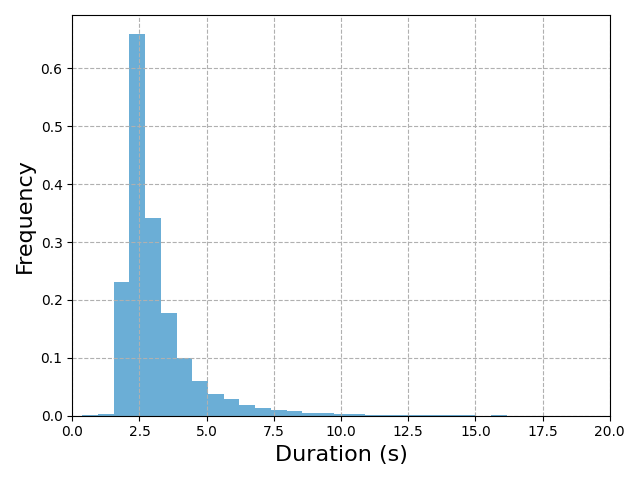}
	\caption{Video duration distribution.}
	\label{Figure6}
\end{figure*}

Figure \ref{Figure6} presents the video duration distribution of the MER-Caption dataset. We observe that the majority of samples have durations ranging from 2 to 5 seconds.

\section{MER-UniBench Details}
\label{appendix:dataset_details}
MER-UniBench is a comprehensive evaluation benchmark covering three typical tasks in MER, including fine-grained emotion recognition, basic emotion recognition, and sentiment analysis. Different tasks involve different datasets, and we provide their statistical information in Table \ref{Table13}. In this paper, we intentionally focus on single-person videos, as this allows us to eliminate interference from other speakers and reduce task difficulty. Multi-person MER belongs to another research topic and will be addressed in our future work.

\begin{table}[h]
	\centering
	\caption{Dataset statistics in MER-UniBench. All datasets in our study focus on single-person videos.}
	\label{Table13}
	\scalebox{0.8}{
		\begin{tabular}{c|lcrlll}
			\hline
			& \textbf{Dataset}  & \textbf{Chosen Set}   & \textbf{\# Samples}  & \textbf{Label Description} & \textbf{Data Source} \\
			\hline
			\textbf{Fine-grained Emotion} & OV-MERD+ & All & 532  & \begin{tabular}[l]{@{}l@{}} unfixed categories and \\ diverse number of labels per sample \end{tabular} & movies, TV series \\
			\hline
			\multirow{4}{*}{\textbf{Basic Emotion}} 
			& MER2023            & MER-MULTI & 411   & most likely label among six candidates  & movies, TV series \\
			& MER2024            & MER-SEMI  & 1,169 & most likely label among six candidates  & movies, TV series \\
			& IEMOCAP       & Session5  & 1,241 & most likely label among four candidates  & actor's performance \\
			& MELD    		   & Test      & 2,610 & most likely label among seven candidates & "Friends" TV series \\
			\hline
			\multirow{4}{*}{\textbf{Sentiment Analysis}} 
			& CMU-MOSI       & Test & 686     & sentiment intensity, ranging from [-3, 3] & opinion videos in YouTube \\
			& CMU-MOSEI  & Test & 4,659  	& sentiment intensity, ranging from [-3, 3] & opinion videos in YouTube \\
			& CH-SIMS               & Test & 457    	& sentiment intensity, ranging from [-1, 1] & movies, TV series, and shows \\
			& CH-SIMS v2        & Test & 1,034  	& sentiment intensity, ranging from [-1, 1] & movies, TV series, and shows \\
			\hline
			
		\end{tabular}
	}
\end{table}

			

\textbf{OV-MERD+} is our newly collected dataset, an extended version of the previous OV-MERD \cite{lian2024open}. Unlike traditional datasets, which select a single label from basic emotions, OV-MERD is a fine-grained emotion dataset that allows each sample to have a variable number of emotions, using any emotion not restricted to predefined taxonomies. OV-MERD initially contains 332 samples, and we further expand its dataset size, obtaining OV-MERD+.

\textbf{MER2023} \cite{lian2023mer} and \textbf{MER2024} \cite{lian2024mer} are widely used in Chinese MER research, with MER2024 being an extended version of MER2023. The original data in both datasets comes from movies and TV shows. They use various techniques to segment video clips, ensuring that each clip has only one person, with their speech content being relatively complete. To ensure annotation quality, they hire multiple annotators, each selecting the most likely label from six candidate emotions: \emph{worry}, \emph{happy}, \emph{neutral}, \emph{angry}, \emph{surprised}, and \emph{sad}. The final label is determined through majority voting.

\textbf{IEMOCAP} \cite{busso2008iemocap} is one of the most widely used emotion datasets. It contains five sessions, each with a male and a female actor in a laboratory environment. The dataset includes the following emotion labels: \emph{anger}, \emph{happiness}, \emph{sadness}, \emph{neutral}, \emph{excitement}, \emph{frustration}, \emph{fear}, \emph{surprise}, and \emph{others}. Following previous research \cite{poria2017context}, we choose the last session for testing, and use the first four emotions, and merge \emph{surprise} and \emph{happiness} into \emph{happiness}.

\textbf{MELD} \cite{poria2019meld} is an extension of the text-centered EmotionLines dataset \cite{chen2018emotionlines}, adding audio and video content. The raw data is derived from the Friends TV series. The dataset has seven emotion labels, and each sample is assigned to one of the most likely labels: \emph{anger}, \emph{joy}, \emph{sadness}, \emph{neutral}, \emph{disgust}, \emph{fear}, and \emph{surprise}.

\textbf{CMU-MOSI} \cite{zadeh2017tensor} and \textbf{CMU-MOSEI} \cite{zadeh2018multimodal} consist of opinion videos collected from online platforms. CMU-MOSEI is an extended version of CMU-MOSI, with more samples and a wider range of topics. In these datasets, each sample is labeled with a sentiment intensity score ranging from -3 to +3, where -3 represents extremely negative emotion and +3 represents extremely positive emotion.

\textbf{CH-SIMS} \cite{yu2020ch} and \textbf{CH-SIMS v2} \cite{liu2022make} differ from the English-centered CMU-MOSI and CMU-MOSEI by focusing on emotions within the Chinese culture. The original data comes from movies, TV series, and shows. Similar to CMU-MOSI and CMU-MOSEI, these datasets also annotate sentiment intensity, but with a different range $[-1, 1]$, where -1 represents extremely negative emotion and +1 represents extremely positive emotion.

\section{Emotion Wheel}
\label{appendix:emotion_wheel}
Since there is no universal definition of the emotion wheel, we follow previous work \cite{lian2024open} and use five emotion wheels in this paper.

\begin{figure*}[h]
	\begin{center}
		\subfigure[\scriptsize{W1}]{
			\label{Figure7-1}
			\centering
			\includegraphics[width=0.26\linewidth, trim=0 0 0 0]{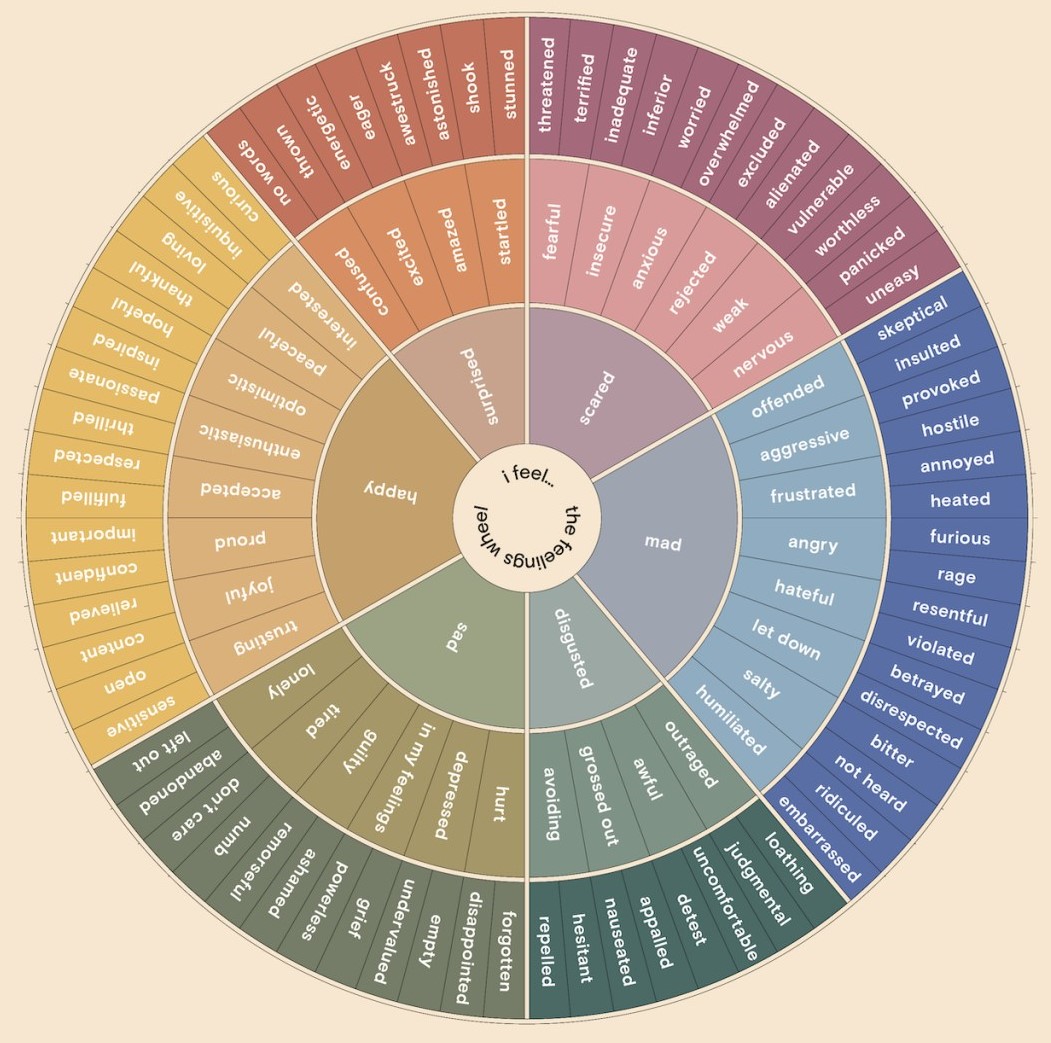}
		}
		\subfigure[\scriptsize{W2}]{
			\label{Figure7-2}
			\centering
			\includegraphics[width=0.26\linewidth, trim=0 0 0 0]{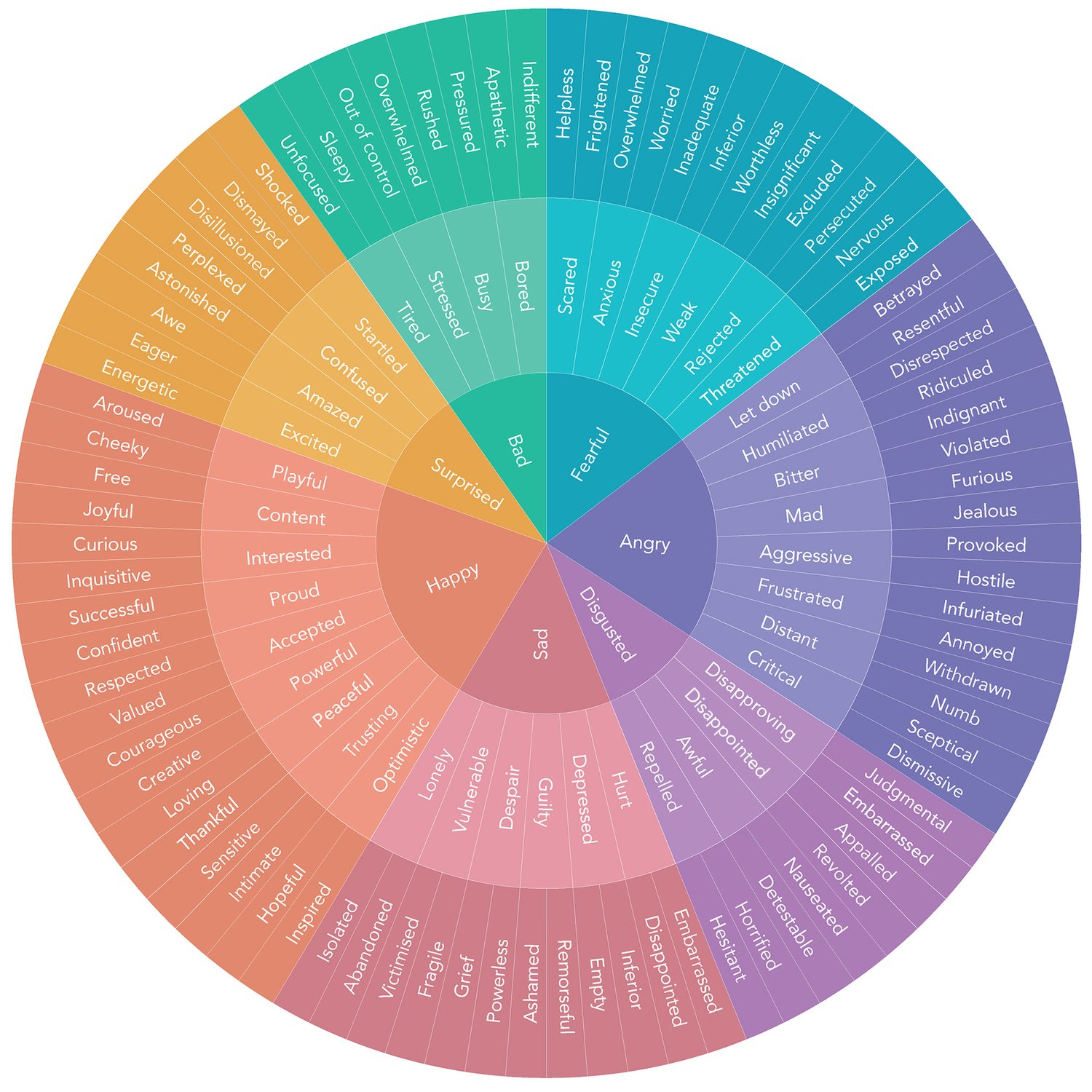}
		}
		\subfigure[\scriptsize{W3}]{
			\label{Figure7-3}
			\centering
			\includegraphics[width=0.26\linewidth, trim=0 0 0 0]{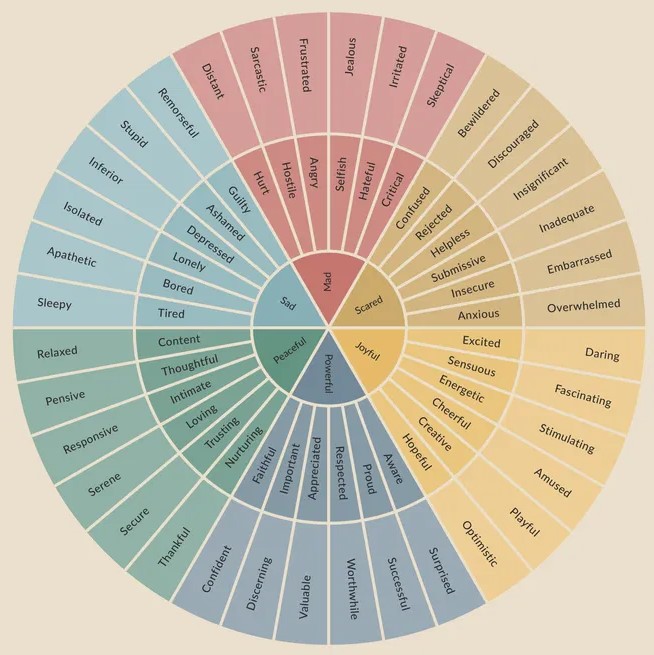}
		}
		
		\subfigure[\scriptsize{W4}]{
			\label{Figure7-4}
			\centering
			\includegraphics[width=0.26\linewidth, trim=0 0 0 0]{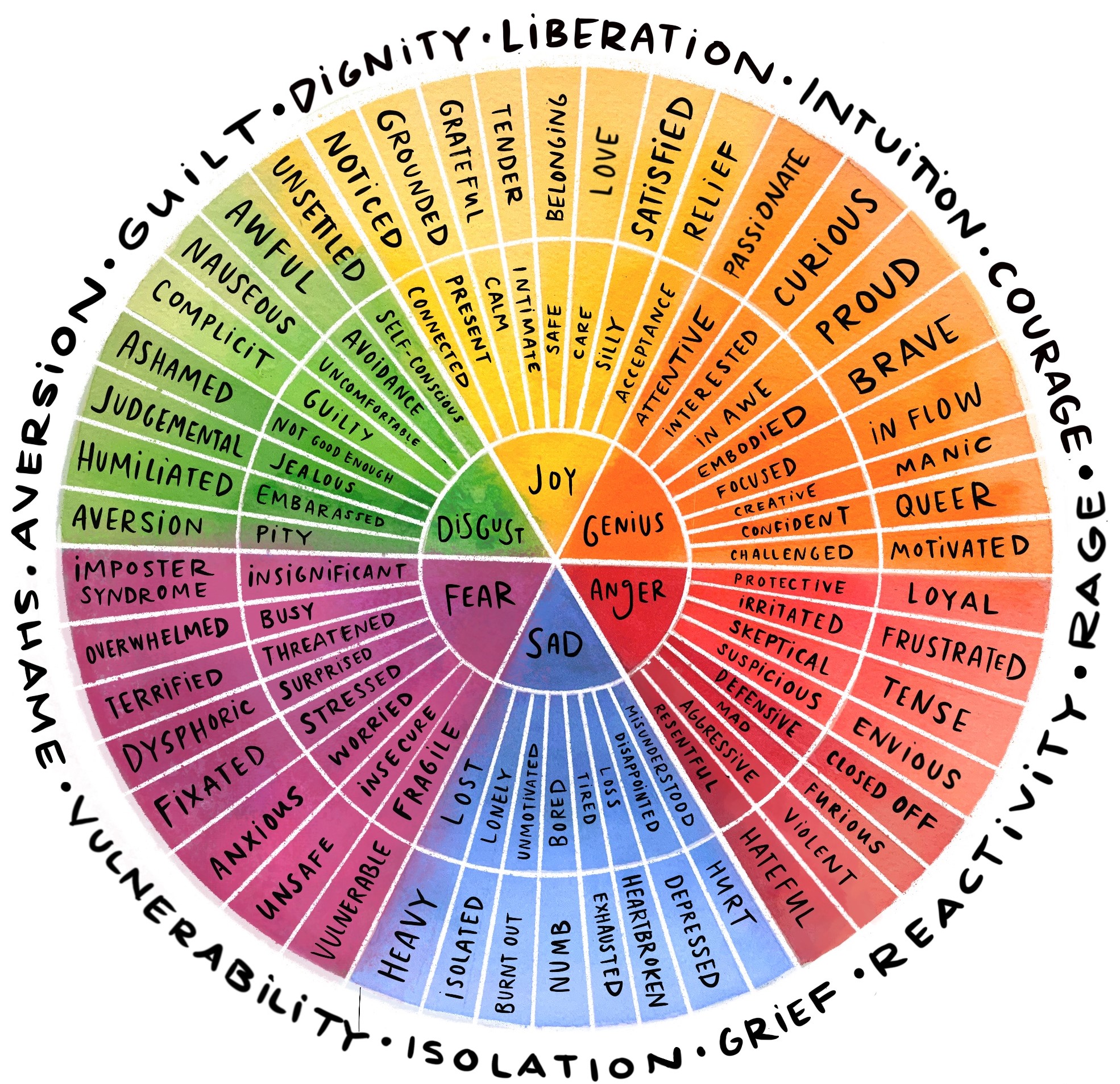}
		}
		\subfigure[\scriptsize{W5}]{
			\label{Figure7-5}
			\centering
			\includegraphics[width=0.26\linewidth, trim=0 0 0 0]{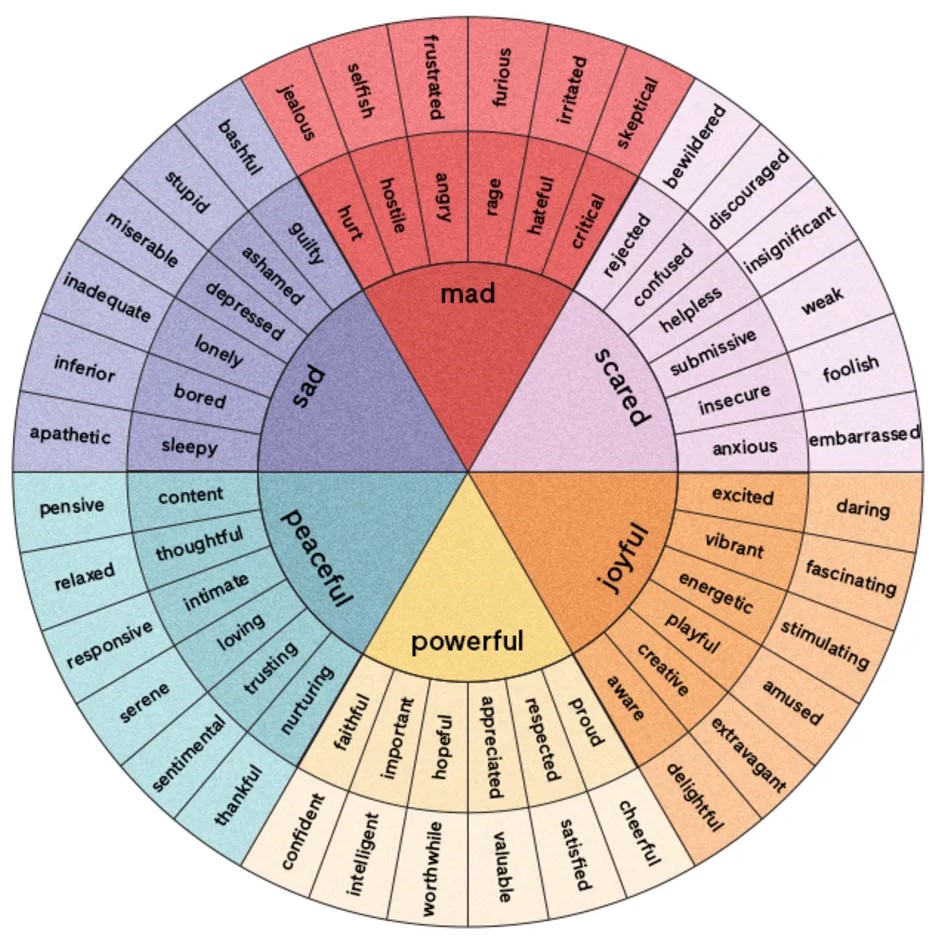}
		}
	\end{center}
	\caption{\textbf{Emotion wheel}. We use five emotion wheels, all of which are derived from previous research \cite{lian2024open}.}
	\label{Figure7}
\end{figure*}

\section{Main Results}
\label{appendix:complte_results}
Table \ref{Table14} reports the complete results, with several metrics for each dataset, and the primary metrics are highlighted in gray. In the last column, we report the average value of the primary metrics. These results verify the effectiveness of our AffectGPT in multimodal emotion understanding.

\begin{table}[h]
	\centering
	\caption{\textbf{Main results}. In this table, ``A'', ``V'', and ``T'' represent audio, video, and text, indicating the input information used by each MLLM during inference. The gray-highlighted columns represent the primary metric for each dataset, while the ``Mean'' column reports the average score of the primary metrics across all datasets.}
	\label{Table14}
	\scalebox{0.8}{
		\renewcommand\tabcolsep{8.3pt}
		\begin{tabular}{lccc|>{\columncolor{lightgray}}c|>{\columncolor{lightgray}}c|>{\columncolor{lightgray}}c|>{\columncolor{lightgray}}c|>{\columncolor{lightgray}}cc|>{\columncolor{lightgray}}cc}
			\hline
			&\multirow{2}{*}{A} &\multirow{2}{*}{V} &\multirow{2}{*}{T}
			&\textbf{MER2023}  
			&\textbf{MER2024} 
			&\textbf{MELD}
			&\textbf{IEMOCAP} 
			&\multicolumn{2}{c|}{\textbf{CMU-MOSI}} 
			&\multicolumn{2}{c}{\textbf{CMU-MOSEI}} \\
			& & & &HIT($\uparrow$) &HIT($\uparrow$) &HIT($\uparrow$) &HIT($\uparrow$) &WAF($\uparrow$) &ACC($\uparrow$)  &WAF($\uparrow$) &ACC($\uparrow$) \\
			\hline
			Otter         &$\times$ &$\surd$  &$\surd$ & 16.41 & 14.65 & 22.57 & 29.08 & 52.89 & 54.27 & 50.44 & 50.77\\
OneLLM        &$\surd$  &$\times$ &$\surd$ & 25.52 & 17.21 & 28.32 & 33.44 & 64.01 & 64.48 & 54.09 & 54.18\\
Video-LLaVA   &$\times$ &$\surd$  &$\surd$ & 36.93 & 30.25 & 30.73 & 38.95 & 56.37 & 57.62 & 61.64 & 64.20\\
SECap         &$\surd$  &$\times$ &$\surd$ & 40.95 & 52.46 & 25.56 & 36.92 & 55.76 & 56.71 & 54.18 & 53.85\\
PandaGPT      &$\surd$  &$\times$ &$\surd$ & 33.57 & 39.04 & 31.91 & 36.55 & 66.06 & 65.85 & 61.33 & 60.73\\
Qwen-Audio    &$\surd$  &$\times$ &$\surd$ & 41.85 & 31.61 &\textcolor[rgb]{1,0.72,0.72}{\textbf{49.09}}& 35.47 & 70.09 & 71.49 & 46.90 & 51.16\\
PandaGPT      &$\times$ &$\surd$  &$\surd$ & 39.13 & 47.16 & 38.33 & 47.21 & 58.50 & 60.21 & 64.25 & 65.55\\
Video-ChatGPT &$\times$ &$\surd$  &$\surd$ & 44.86 & 46.80 & 37.33 &\textcolor[rgb]{1,0.36,0.36}{\textbf{56.83}}& 54.42 & 57.77 & 63.12 & 65.66\\
VideoChat2    &$\times$ &$\surd$  &$\surd$ & 33.67 & 54.50 & 36.64 & 48.70 & 66.84 & 67.23 & 54.32 & 54.82\\
PandaGPT      &$\surd$  &$\surd$  &$\surd$ & 40.21 & 51.89 & 37.88 & 44.04 & 61.92 & 62.80 & 67.61 &\textcolor[rgb]{1,0.72,0.72}{\textbf{68.82}}\\
LLaMA-VID     &$\times$ &$\surd$  &$\surd$ & 50.72 & 57.60 & 42.75 & 46.02 & 61.78 & 62.65 & 63.89 & 66.21\\
VideoChat     &$\times$ &$\surd$  &$\surd$ & 48.73 & 57.30 & 41.11 & 48.38 & 65.13 & 65.09 & 63.61 & 63.02\\
SALMONN       &$\surd$  &$\times$ &$\surd$ & 55.53 & 45.38 & 45.62 & 46.84 &\textcolor[rgb]{1,0.36,0.36}{\textbf{81.00}}&\textcolor[rgb]{1,0.36,0.36}{\textbf{81.25}}& 67.03 & 66.90\\
Chat-UniVi    &$\times$ &$\surd$  &$\surd$ &\textcolor[rgb]{1,0.72,0.72}{\textbf{57.62}}&\textcolor[rgb]{1,0.72,0.72}{\textbf{65.67}}& 45.61 & 52.37 & 54.53 & 57.62 & 63.18 & 67.47\\
mPLUG-Owl     &$\times$ &$\surd$  &$\surd$ & 56.86 & 59.89 &\textcolor[rgb]{1,0.36,0.36}{\textbf{49.11}}&\textcolor[rgb]{1,0.72,0.72}{\textbf{55.54}}&\textcolor[rgb]{1,0.72,0.72}{\textbf{72.40}}&\textcolor[rgb]{1,0.72,0.72}{\textbf{72.26}}&\textcolor[rgb]{1,0.36,0.36}{\textbf{72.91}}&\textcolor[rgb]{1,0.36,0.36}{\textbf{73.17}}\\
Emotion-LLaMA &$\surd$  &$\surd$  &$\surd$  &\textcolor[rgb]{1,0.36,0.36}{\textbf{59.38}}&\textcolor[rgb]{1,0.36,0.36}{\textbf{73.62}}& 46.76 & 55.47 & 66.13 & 66.31 &\textcolor[rgb]{1,0.72,0.72}{\textbf{67.66}}& 67.25 \\
\textbf{AffectGPT}     &$\surd$ &$\surd$ &$\surd$ &\textcolor[rgb]{1,0,0}{\textbf{78.54}}&\textcolor[rgb]{1,0,0}{\textbf{78.80}}&\textcolor[rgb]{1,0,0}{\textbf{55.65}}&\textcolor[rgb]{1,0,0}{\textbf{60.54}}&\textcolor[rgb]{1,0,0}{\textbf{81.30}}&\textcolor[rgb]{1,0,0}{\textbf{81.25}}&\textcolor[rgb]{1,0,0}{\textbf{80.90}}&\textcolor[rgb]{1,0,0}{\textbf{80.68}}\\

			\hline
		\end{tabular}
	}
	\scalebox{0.8}{
		\renewcommand\tabcolsep{9.3pt}
		\begin{tabular}{lccc|>{\columncolor{lightgray}}cc|>{\columncolor{lightgray}}cc|>{\columncolor{lightgray}}ccc|c}
			\hline
			&\multirow{2}{*}{A} &\multirow{2}{*}{V} &\multirow{2}{*}{T}
			&\multicolumn{2}{c|}{\textbf{CH-SIMS}} 
			&\multicolumn{2}{c|}{\textbf{CH-SIMS v2}}
			&\multicolumn{3}{c|}{\textbf{OV-MERD+}}
			&\multirow{2}{*}{\textbf{Mean}} \\
			& & & &WAF($\uparrow$) &ACC($\uparrow$) &WAF($\uparrow$) &ACC($\uparrow$) &$\mbox{F}_{\mbox{s}}$($\uparrow$) &$\mbox{Precision}_{\mbox{s}}$($\uparrow$) &$\mbox{Recall}_{\mbox{s}}$($\uparrow$) & \\
			\hline
			Otter         &$\times$ &$\surd$  &$\surd$ & 57.56 & 60.57 & 53.12 & 56.20 & 16.63 & 17.67 & 15.74 & 34.82\\
OneLLM        &$\surd$  &$\times$ &$\surd$ & 63.39 & 63.92 & 61.98 & 62.46 & 22.25 & 24.49 & 20.41 & 41.14\\
Video-LLaVA   &$\times$ &$\surd$  &$\surd$ & 53.28 & 54.64 & 57.45 & 59.28 & 34.00 & 36.48 & 31.86 & 44.40\\
SECap         &$\surd$  &$\times$ &$\surd$ & 59.51 & 62.89 & 57.41 & 60.92 & 36.97 & 43.51 & 32.17 & 46.64\\
PandaGPT      &$\surd$  &$\times$ &$\surd$ & 62.93 & 62.37 & 58.88 & 58.84 & 31.33 & 33.08 & 29.77 & 46.84\\
Qwen-Audio    &$\surd$  &$\times$ &$\surd$ & 70.73 &\textcolor[rgb]{1,0.72,0.72}{\textbf{73.45}}& 65.26 & 68.17 & 32.36 & 38.52 & 27.91 & 49.26\\
PandaGPT      &$\times$ &$\surd$  &$\surd$ & 62.07 & 61.60 & 65.25 & 65.31 & 35.07 & 37.86 & 32.67 & 50.77\\
Video-ChatGPT &$\times$ &$\surd$  &$\surd$ & 64.82 & 64.43 & 65.80 & 66.85 & 39.80 & 43.12 & 36.97 & 52.64\\
VideoChat2    &$\times$ &$\surd$  &$\surd$ & 69.49 & 69.59 & 70.66 & 71.13 & 39.21 & 42.85 & 36.16 & 52.67\\
PandaGPT      &$\surd$  &$\surd$  &$\surd$ & 68.38 & 67.78 & 67.23 & 67.40 & 37.12 & 39.64 & 34.91 & 52.92\\
LLaMA-VID     &$\times$ &$\surd$  &$\surd$ & 69.35 & 68.81 & 67.48 & 67.73 & 45.01 & 46.83 & 43.32 & 56.07\\
VideoChat     &$\times$ &$\surd$  &$\surd$ & 69.52 & 69.33 & 72.14 & 72.12 & 44.52 & 44.55 & 44.49 & 56.71\\
SALMONN       &$\surd$  &$\times$ &$\surd$ & 68.69 & 69.85 & 65.93 & 67.07 & 45.00 & 43.57 & 46.61 & 57.89\\
Chat-UniVi    &$\times$ &$\surd$  &$\surd$ & 68.15 & 67.78 & 66.36 & 67.18 & 48.00 &\textcolor[rgb]{1,0.72,0.72}{\textbf{48.20}}& 47.81 & 57.94\\
mPLUG-Owl     &$\times$ &$\surd$  &$\surd$ &\textcolor[rgb]{1,0.72,0.72}{\textbf{72.13}}& 71.65 &\textcolor[rgb]{1,0.72,0.72}{\textbf{75.00}}&\textcolor[rgb]{1,0.72,0.72}{\textbf{74.97}}&\textcolor[rgb]{1,0.72,0.72}{\textbf{48.18}}& 47.91 &\textcolor[rgb]{1,0.72,0.72}{\textbf{48.47}}&\textcolor[rgb]{1,0.72,0.72}{\textbf{62.45}}\\
Emotion-LLaMA &$\surd$  &$\surd$  &$\surd$ &\textcolor[rgb]{1,0.36,0.36}{\textbf{78.32}}&\textcolor[rgb]{1,0.36,0.36}{\textbf{78.61}}&\textcolor[rgb]{1,0.36,0.36}{\textbf{77.23}}&\textcolor[rgb]{1,0.36,0.36}{\textbf{77.39}}&\textcolor[rgb]{1,0.36,0.36}{\textbf{52.97}}&\textcolor[rgb]{1,0.36,0.36}{\textbf{54.85}}&\textcolor[rgb]{1,0.36,0.36}{\textbf{51.22}}&\textcolor[rgb]{1,0.36,0.36}{\textbf{64.17}}\\
\textbf{AffectGPT}     &$\surd$ &$\surd$ &$\surd$  &\textcolor[rgb]{1,0,0}{\textbf{88.49}}&\textcolor[rgb]{1,0,0}{\textbf{88.40}}&\textcolor[rgb]{1,0,0}{\textbf{86.18}}&\textcolor[rgb]{1,0,0}{\textbf{86.17}}&\textcolor[rgb]{1,0,0}{\textbf{62.52}}&\textcolor[rgb]{1,0,0}{\textbf{62.21}}&\textcolor[rgb]{1,0,0}{\textbf{63.00}}&\textcolor[rgb]{1,0,0}{\textbf{74.77}}\\

			\hline
		\end{tabular}
	}
\end{table}

\section{Ablation Study on MER-Caption}
\label{appendix:ablation_study_dataset}
Table \ref{Table3-complete} compares the performance across different datasets. To ensure a fair comparison, we keep the model architecture and experimental setup unchanged, only altering the training dataset. Experimental results in Table \ref{Table3-complete} demonstrate the effectiveness of our MER-Caption dataset for emotion understanding. It addresses the issue of existing datasets either giving insufficient attention to emotion tasks or lacking high-quality emotion descriptions.
\begin{table*}[h]
	\centering
	\renewcommand\tabcolsep{3.6pt}
	\caption{Dataset comparison.}
	\label{Table3-complete}
	\scalebox{0.8}{
		\begin{tabular}{clc|ccccccccc|c}
			\hline
			& \textbf{Dataset} & \textbf{Filtering}
			& \textbf{MER2023}
			& \textbf{MER2024}
			& \textbf{MELD}
			& \textbf{IEMOCAP}
			& \begin{tabular}[c]{@{}c@{}}\textbf{CMU-}\\\textbf{MOSI}\end{tabular}
			& \begin{tabular}[c]{@{}c@{}}\textbf{CMU-}\\\textbf{MOSEI}\end{tabular}
			& \begin{tabular}[c]{@{}c@{}}\textbf{CH-}\\\textbf{SIMS}\end{tabular}
			& \begin{tabular}[c|]{@{}c@{}}\textbf{CH-}\\\textbf{SIMS v2}\end{tabular}
			& \begin{tabular}[c]{@{}c@{}}\textbf{OV-}\\\textbf{MERD+}\end{tabular}
			& \textbf{Mean} \\
			\hline	
			\multirow{8}{*}{\begin{tabular}[c|]{@{}c@{}}\textbf{General}\\\textbf{Instruction}\end{tabular}} 
			& \multirow{2}{*}{MiniGPT4} 
			& $\times$ & 11.56 & 12.91 & 18.89 & 16.06 & 53.57 & 45.98 & 57.66 & 55.16 & 13.86 & 31.74 \\
			& & $\surd$ & 17.57 & 16.65 & 22.60 & 30.18 & 52.58 & 56.50 & 52.36 & 51.19 & 20.16 & 35.53 \\
			\cline{2-13}
			& \multirow{2}{*}{VideoChat} 
			& $\times$ & 24.87 & 22.42 & 21.56 & 32.91 & 50.13 & 56.17 & 50.07 & 51.71 & 24.56 & 37.16 \\
			& & $\surd$  & 27.70 & 24.73 & 27.66 & 39.46 & 45.45 & 56.86 & 43.68 & 47.05 & 26.09 & 37.63 \\
			\cline{2-13}
			& \multirow{2}{*}{LLaVA} 
			& $\times$ & 42.21 & 41.54 & 32.97 & 49.96 & 54.48 & 56.42 & 52.04 & 54.80 & 35.75 & 46.69 \\
			& & $\surd$ & 41.56 & 42.30 & 32.61 & 46.21 & 52.82 & 57.72 & 52.78 & 53.44 & 36.96 & 46.27 \\
			\cline{2-13}
			& \multirow{2}{*}{WavCaps} 
			& $\times$ & 5.75 & 7.71 & 4.35 & 4.99 & 45.59 & 22.76 & 53.04 & 45.68 & 4.95 & 21.65 \\
			& & $\surd$  & 23.72 & 26.97 & 23.39 & 27.30 & 54.67 & 49.54 & 58.12 & 55.93 & 21.57 & 37.91 \\
			\hline
			\multirow{6}{*}{\begin{tabular}[c|]{@{}c@{}}\textbf{Emotion}\\\textbf{Description}\end{tabular}} 
			& EmoVIT      & -- & 39.31 & 50.24 & 32.36 & 48.24 & 53.40 & 61.53 & 69.72 & 66.53 & 38.09 & 51.05 \\
			& MAFW        & -- & 52.67 & 55.99 & 40.85 & 57.60 & 66.11 & 62.27 & 75.20 & 70.02 & 42.75 & 58.16 \\
			& MERR-Coarse & -- & 35.34 & 36.60 & 29.37 & 36.94 & 65.10 & 63.27 & 75.12 & 73.76 & 33.18 & 49.85 \\
			& MERR-Fine   & -- & 69.00 & 72.84 & 47.38 & 54.49 & 66.21 & 60.03 & 79.90 & 78.54 & 52.56 & 64.55 \\
			& MER-Caption & -- & 72.12 & 74.21 & 54.69 & 56.41 & 75.10 & 70.97 & 80.21 & 78.06 & 58.42 & 68.91 \\
			& MER-Caption+ & --  & \textcolor[rgb]{1,0,0}{\textbf{78.54}}& \textcolor[rgb]{1,0,0}{\textbf{78.80}}& \textcolor[rgb]{1,0,0}{\textbf{55.65}} & \textcolor[rgb]{1,0,0}{\textbf{60.54}} & \textcolor[rgb]{1,0,0}{\textbf{81.30}} & \textcolor[rgb]{1,0,0}{\textbf{80.90}} & \textcolor[rgb]{1,0,0}{\textbf{88.49}} & \textcolor[rgb]{1,0,0}{\textbf{86.18}} & \textcolor[rgb]{1,0,0}{\textbf{62.52}} & \textcolor[rgb]{1,0,0}{\textbf{74.77}} \\
			\hline
		\end{tabular}
	}
\end{table*}

\clearpage

\section{Impact of Sampling Frames in Video Branch}
\begin{figure*}[h]
	\begin{center}
		\subfigure[{Face-only Input}]{
			\label{Figure27-1}
			\centering
			\includegraphics[width=0.36\linewidth, trim=0 0 0 0]{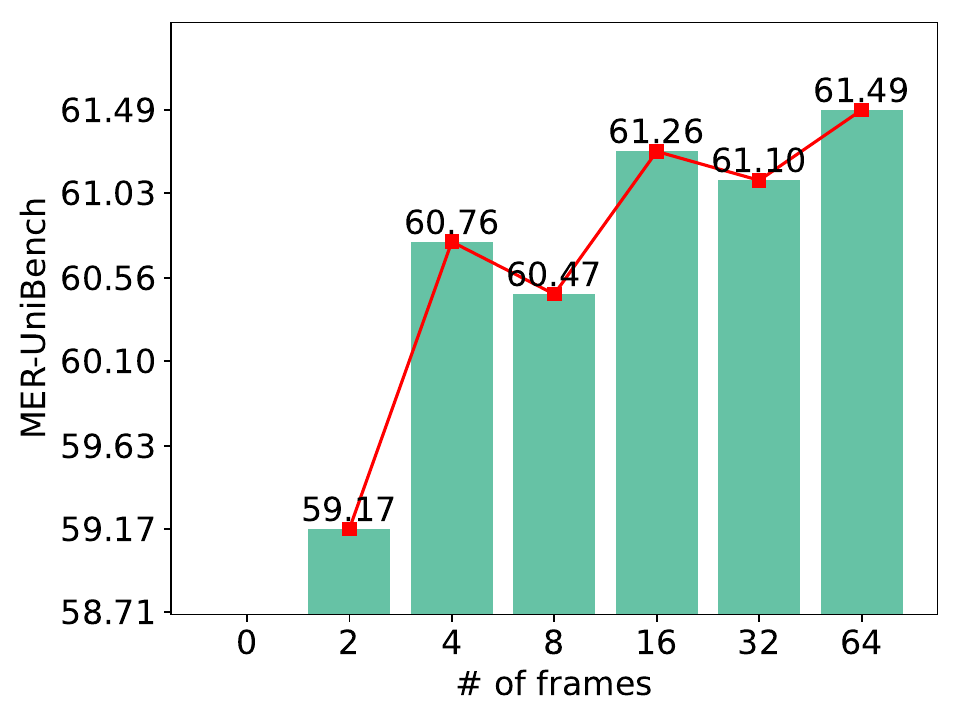}
		}
		\subfigure[{Face-text Input}]{
			\label{Figure27-2}
			\centering
			\includegraphics[width=0.36\linewidth, trim=0 0 0 0]{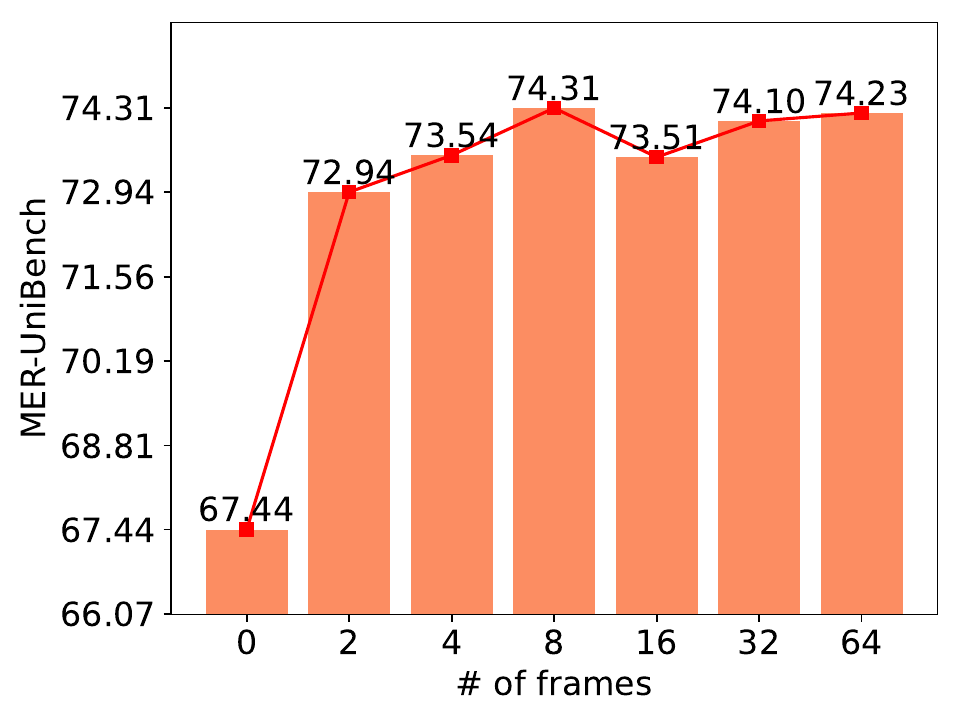}
		}
	\end{center}
	\caption{Impact of sampling frames.}
	\label{Figure27}
\end{figure*}

This paper defaults to sampling 8 frames per video. But if we change the number of sampled frames, will it significantly impact performance? To answer this, we conducted additional experiments in this section. Specifically, we compare two types of inputs: (1) face-only and (2) face-text combinations, and evaluate model performance across different sampling frame counts, ranging from 2 to 64. In Figure \ref{Figure27}, we observe that using too few frames (e.g., fewer than 2) results in a noticeable decline in performance, indicating that insufficient frames lead to information loss. However, further increasing the number of sampling frames (e.g., more than 16) does not yield significant performance improvements. This can be attributed to the fact that MER tasks typically use short-duration videos with relatively stable facial expressions.

\end{document}